%% file: ms.tex
\documentclass[12pt,preprint]{aastex}

\shorttitle{STELLAR ROTATION IN M35}
\shortauthors{Meibom et al.}

\begin{document}

\title{STELLAR ROTATION IN M35: MASS-PERIOD RELATIONS, SPIN-DOWN RATES,
AND GYROCHRONOLOGY\altaffilmark{1}}

\author{S{\o}ren Meibom\altaffilmark{2,3,4} and Robert D. Mathieu
\altaffilmark{4}}
\affil{Department of Astronomy, University of Wisconsin - Madison, Madison,
WI, 53706, USA}

\author{Keivan G. Stassun\altaffilmark{4}}
\affil{Physics and Astronomy Department, Vanderbilt University, Nashville, TN - 32735, USA}

\altaffiltext{1}{WIYN Open Cluster Study. XXXIII.}
\altaffiltext{2}{{\it smeibom@cfa.harvard.edu}}
\altaffiltext{3}{{Now at Harvard-Smithsonian Center for Astrophysics, Cambridge, MA, 02138, USA}}
\altaffiltext{4}{Visiting Astronomer, Kitt Peak National Observatory,
National Optical Astronomy Observatory, which is operated by the
Association of Universities for Research in Astronomy, Inc. (AURA)
under cooperative agreement with the National Science Foundation.}

% *****************************************************************************

\begin{abstract} \label{abs}

We present the results of a 5 month photometric time-series survey
for stellar rotation over a $40\arcmin \times 40\arcmin$ field centered
on the 150\,Myr open cluster M35. We report rotation periods for 441 stars
within this field and determine their cluster membership and binarity
based on a decade-long radial-velocity survey, proper-motion measurements,
and multi-band photometric observations. We find that 310 of the stars
with measured rotation periods are late-type members of M35. The distribution
of rotation periods for cluster members span more than two orders of
magnitude from $\sim 0.1 - 15$ days, not constrained by the sampling
frequency and the time-span of the survey. With an age between the zero-age
main-sequence and the Hyades, and with $\sim$6 times more rotation
periods than measured in the Pleiades, M35 permit detailed
studies of early rotational evolution of late-type stars. Nearly 80\%
of the 310 rotators lie on two distinct sequences in the color-period
plane, and define clear relations between stellar rotation period and color
(mass). The M35 color-period diagram enables us to determine timescales
for the transition between the two rotational states, of $\sim$60\,Myr
and $\sim$140\,Myr for G and K dwarfs, respectively. These timescales
are inversely related to the mass of the convective envelope, and offer
constraints on the rates of internal and external angular momentum
transport and of the evolution of stellar dynamos. A comparison to the
Hyades, confirm the Skumanich (1972) spindown-dependence for G dwarfs
on one rotational state, but suggest that K dwarfs spin down more slowly.
The locations of the rotational sequences in the M35 color-period diagram
support the use of rotational isochrones to determine ages for coeval
stellar populations. We use such gyrochronology to determine ``gyro-ages''
for M35 from 134\,Myr to 161\,Myr. We use the M35 data to evaluate new color
dependencies for the rotational isochrones.

\end{abstract}


% *****************************************************************************

\section{INTRODUCTION}
\label{intro}

Observations of coeval populations of late-type stars younger
than the Hyades have revealed that they rotate with periods
ranging over two orders of magnitude - from near breakup to
periods similar to the Sun. Understanding why some stars
deplete their angular momentum faster than others, which
physical processes are at work, when, and to what extent,
is a primary mandate for stellar evolution research.

The discovery from photometric and spectroscopic measurements
in the Pleiades of sub 1-day rotation periods for K dwarfs
\citep{av81,va82,mav82,sjw83} challenged prior
understanding of the early angular momentum evolution for late-type stars,
and sparked a renewed interest in the topic. Observations of mainly projected
rotation velocities ($v \sin(i)$) of late-type stars in $\alpha$ Persei
(50 Myr; \citet{shb+85,shj89}), the Pleiades (100 Myr;
\citet{sjw83,shs+84,bmm84,sh87,ssh+93,tsp+00}), and the Hyades (625 Myr;
\citet{soderblom82,bmm84,rtl+87}), and photometric studies of the Hyades
\citep{ltr+84,rtl+87}, confirmed the coexistence of slowly and rapidly
rotating stars in $\alpha$ Persei and the Pleiades, but found an absence
of rapid rotators in the Hyades. Furthermore, the rapid rotators in the
youngest clusters were found primarily among K and M dwarfs and not among
G dwarfs.

The emerging evidence for an age- and mass-dependence of rapid rotation
prompted new ideas about the rotational evolution of late-type stars.
For example the suggestion of epochs of decoupling and recoupling of the
stellar core and envelope \citep[e.g.][]{shs+84,ssm+93,jc93}. The idea of
decoupling - allowing the more shallow convective envelope of G dwarfs
to spin down faster than the envelopes in K and M dwarfs - had developed
in parallel in models of stellar rotation \citep[e.g.][]{es81,pkd90,mb91,bs96}.
However, the concept of decoupling, if permanent, is in conflict with
helioseismic observations of the sun as a solid-body rotator
\citep{gough82,ddg+84,gdk+91,ekj02}. Furthermore, recoupling - giving
access to the angular momentum reservoir of the faster spinning core -
was suggested by \citet{ssm+93} as being necessary to explain the evolution
beyond the age of the Pleiades of slowly rotating stars \citep{ssm+93}.

Understanding the formation of the rapid rotators is a separate problem.
The fastest spinning stars in the youngest clusters cannot be explained
from Skumanich-style spin-down \citep{skumanich72} of the fastest spinning
T Tauri stars. The rapid rotators can be explained only by introducing
``magnetic saturation'' of the angular momentum loss via stellar winds
\citep{sh87,mb91,bs96,bfa97,kpb+97,spt00}, and by allowing the saturation
threshold to depend on the stellar mass. The physical meaning of
``saturation'' is still unclear.

During the pre main-sequence (PMS) phase, large dispersions and substructure
(bimodalities) in the rotation-period distributions has also been observed
for coeval stellar populations. Here, a popular explanation for coeval rapid
and slow rotators originates from the work of \citet{konigl91} and
\citet{esh+93} on interactions between T Tauri stars and their circumstellar
disks. ``Magnetic disk-locking'' was introduced to provide a means to brake
the spin-up of the central star by transferring angular momentum from the
star to the disk \citep[e.g.][and references therein]{sno+94,najita95}.
Accordingly, rotation-period dispersions (bimodalities) should result if
some stars lose their disks faster than others (e.g. weak vs. classic T
Tauri stars). Whether magnetic disk-braking is a dominant process regulating
stellar rotation during the early PMS remains under debate on both
observational and theoretical grounds.

Recently, taking advantage of results from an increasing number of
photometric monitoring programs of late-type stars in young main-sequence
clusters, \citet{barnes03a} presented an interpretation of his own
and other published rotation period data. Free of the ambiguities of
$v \sin(i)$, \citeauthor{barnes03a} identified, in each coeval stellar
population, separate groups of fast and intermediate/slowly rotating
stars with different dependencies on color (mass). He specifically
proposed that coeval stars fall along two ``rotational sequences'' in
the color vs. rotation period plane. From an analysis of these sequences
and their dependencies on stellar age, \citet{barnes03a} proposed a
framework for connecting internal and external magneto-hydrodynamic
processes to explain the evolution in the observed period distributions,
including bimodalities. This approach combines the ideas of
\citep[e.g.][]{shs+84,ssm+93} of an initial decoupling of the stellar
core and envelope with re-connection of the two zones through a global
dynamo-field at a later and mass-dependent time. It does not (yet)
interface with PMS angular momentum evolution.
Importantly, \citet{barnes03a} also proposed that the age-dependence of
the location of the rotational sequences in the color-period plane,
could be used to measure the age of a stellar population, much like
the sequences in the color-magnitude diagram. \citet{barnes07} further
developed this idea of ``gyrochronology''.
Determining the age of a late-type star from its rotation
and color, had previously been suggested by \citet{kawaler89}.
% based on theoretical models of angular momentum loss.

Interpretation aside, it has become desirable and increasingly
possible to eliminate the ambiguity of projected rotation velocities
($v\sin i$) by measuring photometric rotation periods from light
modulation by spots on the surfaces of young late-type stars.
While more labor and time intensive, photometric measurements of
rotation periods in coeval populations,
promise to reveal important details about dependencies of rotation
on other stellar properties - most obviously mass and age, but
likely also stellar activity, internal/external magnetic configurations,
and binarity.

We present in this paper the results of an extensive time-series
photometric survey for rotation periods and a decade-long spectroscopic
surveys for membership and binarity for late-type stars in the field
of the open cluster M35 (NGC\,2168). The combination of time-series
and multi-band photometry with time-series radial-velocity data enable
us to explore the distribution of rotation periods vs. stellar color
(mass) for {\it bona fide} single and binary members of M35.

M35 is a rich ($\ga$ 2500 stars; \citet{nds01}) northern hemisphere
cluster located $\sim$800-900 pc \citep{nds01,kfr+03} toward the
galactic anti-center ($\alpha_{2000} = 6^{h}~9^{m}$, $\delta_{2000}
= 24\degr~20\arcmin$; $l = 186\fdg59$, $b = 2\fdg19$). The age of
M35 has been estimated to 150\,Myr \citep{vss+02}, 175\,Myr \citep{nds01},
and 180\,Myr \citep{kfr+03}. We adopt an age of 150\,Myr, in agreement
with the most recent age determination based on the isochrone method
\citep{deliyannis08}.
At the distance of M35 the majority of cluster members are confined
to within a $\sim 0.5\degr$ diameter field, facilitating photometric
and spectroscopic observations of a large number of stars through
wide-field CCD imaging and multi-object spectroscopy. Older and much
more populous than the Pleiades, and younger than the Hyades, M35
nicely bridges a gap in the age sequence of star clusters with
comprehensive information about rotation and membership, permitting
a more detailed study of the rotational evolution of late-type stars
beyond the zero-age main-sequence (ZAMS).

We begin in Section~\ref{obs} by describing our time-series photometric
observations, our methods for data-reduction and for photometric period
detection, and the information about cluster membership available from
our spectroscopic survey and the M35 color-magnitude diagram. In
Section~\ref{distr} we present the distribution of rotation periods
in M35, discuss the short- and long-period tails in the context of our
period detection limits, and assess the stability/lifetime of spots or
groups of spots by comparison of our short- and long-term photometric data.
Section~\ref{cpd} introduces the M35 color-period diagram and the
observed dependencies of stellar rotation on mass. In Section~\ref{discuss}
we discuss and interpret the M35 color-period diagram in the context
of present ideas for stellar angular momentum evolution. In particular
we use the diagram to estimate spin-down rates for G and K dwarfs, to
test the time-dependence on rotational evolution from a comparison with
measured rotation periods in the Hyades, determine M35's gyrochronology
age, and to evaluate the best functional representation of the color-period
dependence in the M35 data. Section~\ref{conclusions} summarizes and
presents our conclusions.

% *****************************************************************************

\section{OBSERVATIONS, DATA REDUCTION, AND METHODS} \label{obs}

\subsection{Time-Series Photometric Observations}     \label{programs}

We photometrically surveyed stars in a region approximately
$40\arcmin \times 40\arcmin$ centered on the open cluster M35
over a timespan of 143 days. The photometric data were
obtained in the Johnson V-band with the WIYN \footnote{The WIYN
Observatories are joint facilities of the University of Wisconsin-Madison,
Indiana University, Yale University, and the National Optical Astronomy
Observatories.} 0.9m telescope \footnote{The 0.9m telescope is operated
by WIYN Inc. on behalf of a Consortium of ten partner Universities
and Organizations (see http://www.noao.edu/0.9m/general.html)} on
Kitt Peak equipped with a $2k \times 2k$ CCD
camera. The field of view of this instrument is $20.5\arcmin
\times 20.5\arcmin$ and observations where obtained over a $2
\times 2$ mosaic.

The complete dataset presented is composed of images from
high-frequency (approximately once per hour for 5-6 hours per night)
time-series photometric observations over 16 full nights from 2-17 December
2002, complemented with a queue-scheduled observing program over
143 nights from 22 October 2002 to 11 March 2003, obtaining one
image per night interrupted only by bad weather and scheduled
instrument changes. The result is a database of differential V-band
light curves for more than 14,000 stars with $12 \la V \la 19.5$.
The sampling frequency of the December 2002 observations allow
us to detect photometric variability with periods ranging from
less than a day to about 10 days. The long time-span of the
queue-scheduled observations provide data suitable for detecting
periodic variability of up to $\sim$ 75 days, and for testing
the long-term stability of short-period photometric variations.
From this database we derive rotation periods for 441 stars.

\begin{figure}[ht!]
\epsscale{1.0}
\plotone{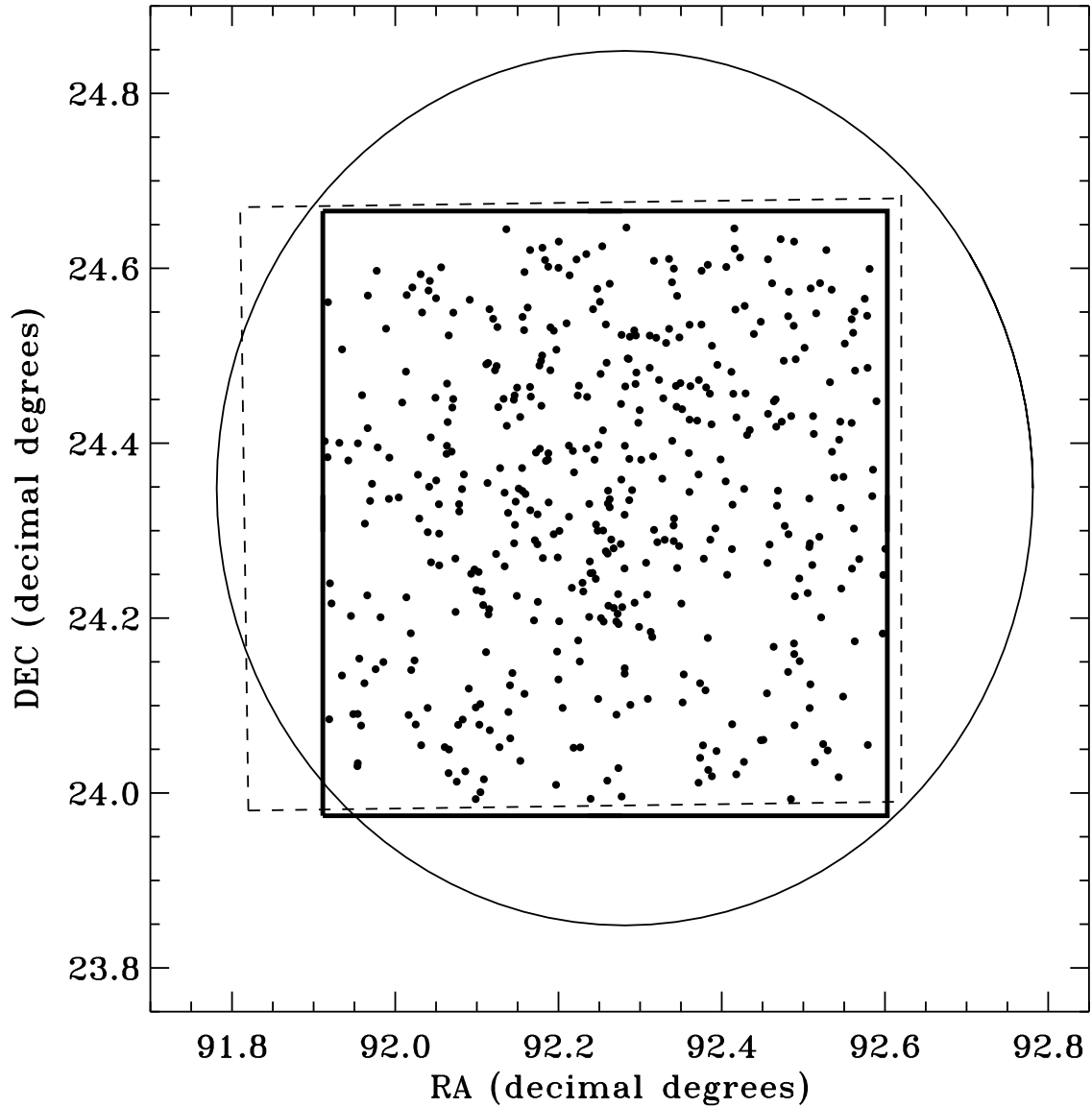}
\caption{The locations and spatial extents of the photometric
and spectroscopic surveys used in this study. The innermost
solid square denotes the $40\arcmin \times 40\arcmin$ region
of our time-series photometric survey. Within it we show the
distribution of the 441 stars with measured rotation periods
(black dots). The dashed rectangle
displays the region of the multi-band photometric study by
\citet{deliyannis08} and the circle represents the 1-degree
diameter field of our spectroscopic survey of M35 \citep{mm05}.
\label{spatial}}
\end{figure}

Figure~\ref{spatial} shows the surveyed region (solid square)
and the spatial distribution of the 441 rotators. The region
is roughly coincident with that of \citet{deliyannis08} in which
they obtained UBVRI CCD photometry for $\sim$19,000 stars (dashed
square). Also shown is the circular target region of our spectroscopic
survey described in Section~\ref{spec} and in \citet{mm05}. The
photometric survey was carried out within the region of the
spectroscopic survey to optimize information about spectroscopic
membership and binarity\footnote{All of these mutually supportive
studies are parts of the WIYN Open Cluster Study (WOCS; \citet{mathieu00}).}.

%\clearpage
%\clearpage

\begin{figure}[ht!]
\epsscale{1.0}
\plotone{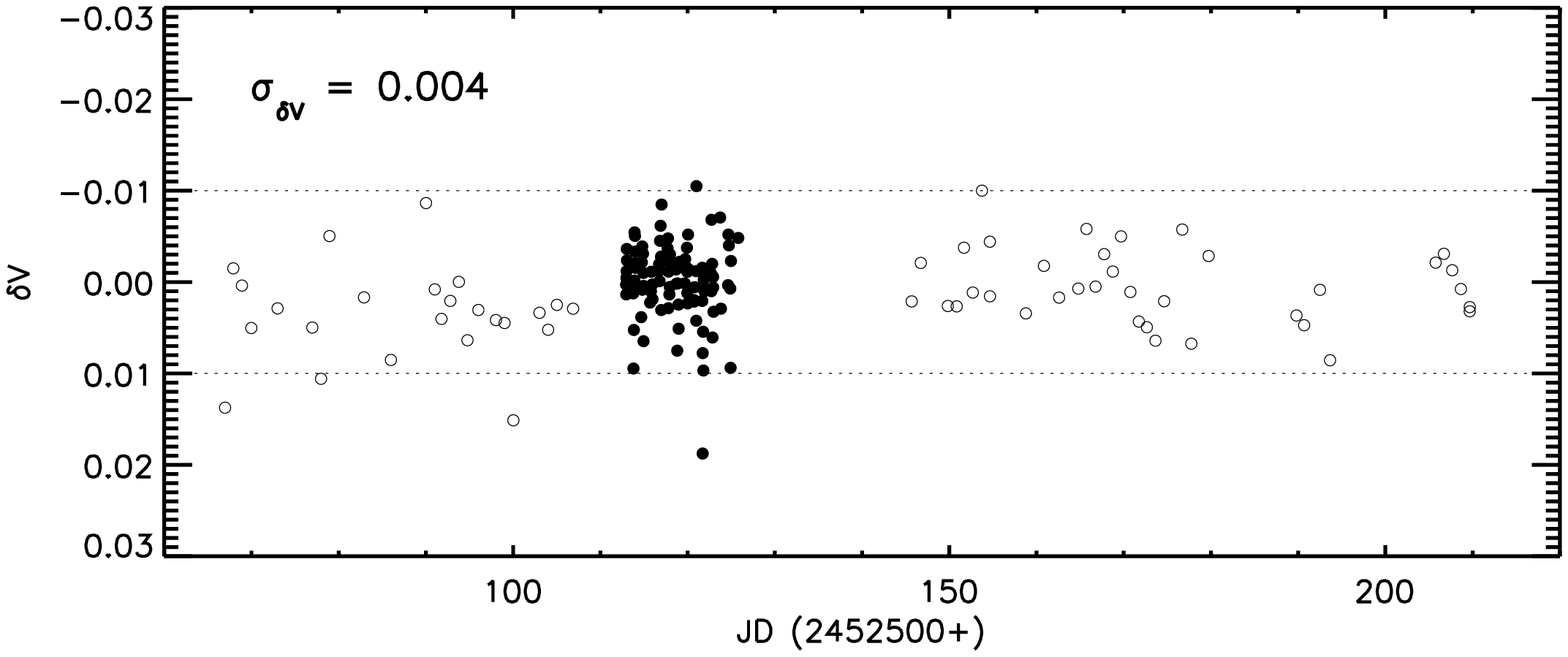}
\caption{Sample time-series data from from our photometric database
for a non-variable $V \simeq$14th magnitude star. Filled symbols
represent measurements from the high-frequency December 2002 observing
run and open symbols represent the low-frequency queue-scheduled
observations. The data span a total of 143 days. The star was observed
in all 157 images of the north-east quadrant of the $2 \times 2$ mosaic.
The standard deviation ($\sigma_{\delta V}$) of the 157 measurements
is $0\fm004$, representative of our best photometric precision. The
horizontal dotted lines denote $\delta V = \pm 0\fm01$.
\label{sampling}}
\end{figure}

Figure~\ref{sampling} displays the time-series data from both
programs for a photometrically non-variable star. Filled symbols
represent the high-frequency observations and open symbols
represent the queue-scheduled observations.

%\clearpage
%\clearpage

% -----------------------------------------------------------------------------

\subsection{Basic Reductions, PSF Photometry, and Light Curves}	\label{psf}

Basic reductions of our CCD frames, identification of stellar sources,
and computations of equatorial coordinates\footnote{We used data from
the STScI Digitized Sky Survey; The Digitized Sky Surveys were produced
at the Space Telescope Science Institute under U.S. Government grant
NAG W-2166.} were done using standard IRAF packages.
Instrumental magnitudes were determined from Point Spread Function
(PSF) photometry using the IRAF DAOPHOT package. The analytical PSF
and a residual lookup table were derived for each frame based on
$\sim$30 evenly distributed isolated stars. The number of measurements
in the light curve of a given
star vary because stars near the edges of individual frames may be
missed due to telescope pointing errors, while bright stars near the
CCD saturation limit and faint stars near the detection threshold may
be excluded on some frames because of variations in seeing, sky brightness,
and sky transparency. To ensure our ability to perform reliable
time-series analysis on stars in our database, we have eliminated
stars that appear on fewer than half of the total number of frames.
The resulting database contains 14022 stars with a minimum of 75
measurements.

\begin{figure}[ht!]
\epsscale{1.0}
\plotone{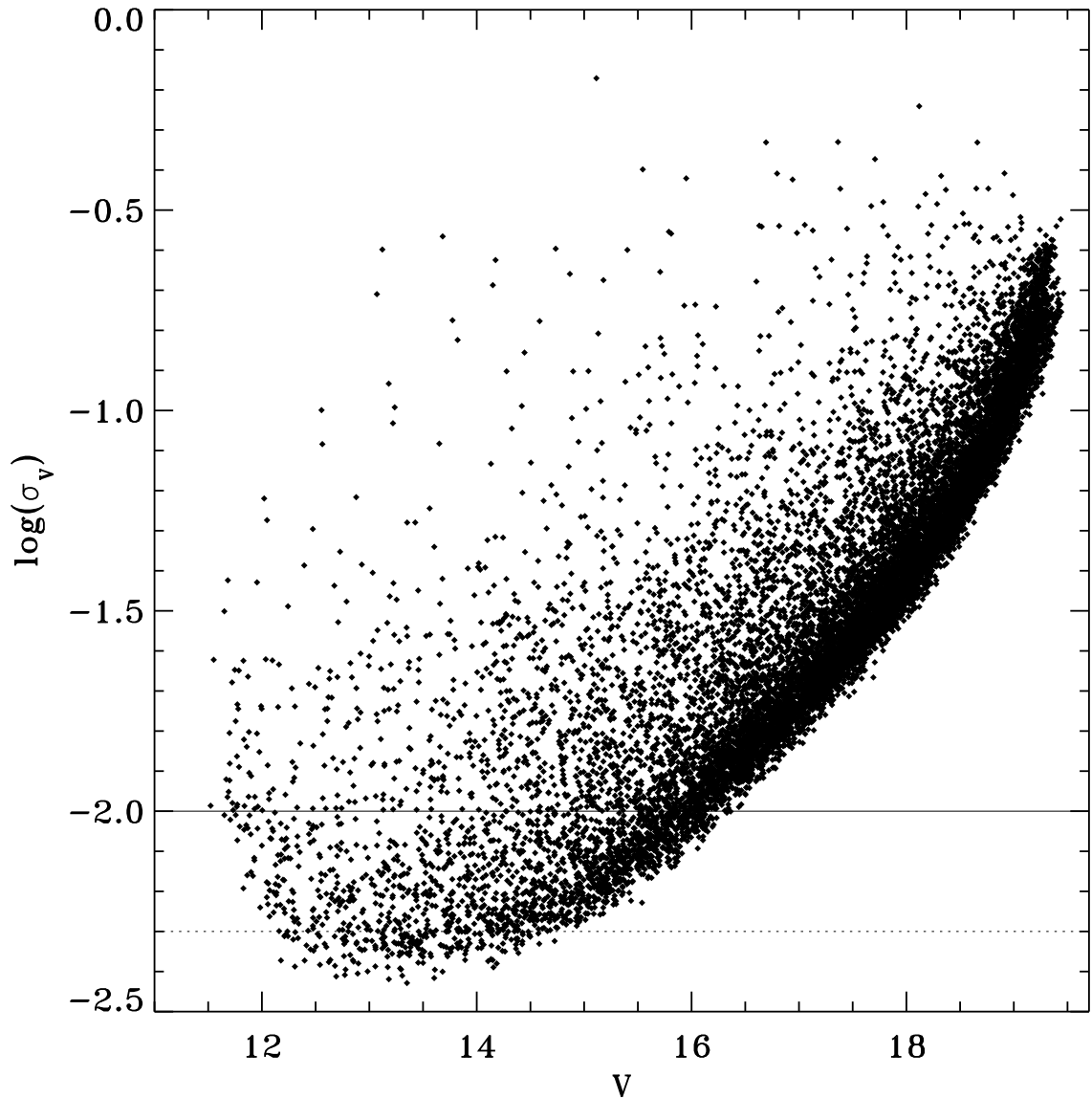}
\caption{The logarithm of the standard deviation of all instrumental
magnitudes as a function of V magnitude for 14022 stars in the field
of M35. The number of measurements for each star range from 75 to 216.
The solid and dashed
horizontal lines represent $\sigma_V$ of 0.01 (1\%) and 0.005 (0.5\%),
respectively. A relative photometric precision of $\sim$ 0.5\% is
obtained for stars with $12\fm0 \la V \la 14.5\fm0$.
\label{m0_sigst}}
\end{figure}

We applied the \citet{honeycutt92} algorithm for differential
CCD photometry to our raw light curves to remove non-stellar
frame-to-frame photometric variations. We favor this technique
for differential photometry because it does not require a
particular set of comparison stars to be chosen {\it a priori},
nor does it require a star to appear in every frame. 
Figure~\ref{sampling} shows the light curve for a  $V \sim$14th
magnitude star. The standard deviation from 157 photometric
measurements is 0\fm004, representative of our photometric
precision at that brightness. Figure~\ref{m0_sigst} shows the
standard deviation of the photometric measurements as a function
of the V magnitude for each star in the field of M35. The relative
photometric precision is $\sim$0.5\% for stars with $12 \la V \la 14.5$.

%\clearpage
%\clearpage

% -----------------------------------------------------------------------------

\subsection{Photometric Period Detection}		\label{periods}

We employed the \citet{scargle82} periodogram analysis to detect
periodic variability in the light curves because of its ability to
handle unevenly sampled data. We searched a grid of 5000 frequencies
corresponding to periods between 0.1 day and 90 days. The lower
search limit was set at a period ensuring critical sampling based
on the Nyquist critical frequency for our high-frequency data
($f_{c} = 1/(2 \delta t)$, where $\delta t$ is the sampling interval
of $\sim$1 hour. The upper limit was set at 90 days because a star
with a 90-day period would complete about 1.5 cycle over the 143
nights of the survey.

A false alarm probability (FAP), the probability that a signal
detected at a certain power level can be produced by statistical
fluctuation, was calculated as the measure of confidence in a
detected period. An analytical expression for estimating a FAP
is given by \citet{scargle82} and \citet{hb86}. However, these
methods are not entirely suitable when applied to time-series
photometric studies of young stars because they only test against
random fluctuations of a purely statistical nature (i.e., measurement
errors) and do not account for correlated fluctuations intrinsic
to the source such as variability on timescales long compared to
the sampling frequency. For young stars our repeated measurements
during a single night are not necessarily independent and uncorrelated.
Consequently, the analytical expressions estimating a FAP will
likely overestimate the significance of any measured periodic
variability. Hence, we performed a two-dispersion Monte Carlo
calculation to estimate the FAP of our detected periods, as per
\citet{hw96} and \citet{smm+99}.
For each star, we generated a set of 100 synthetic
light curves, each consisting of normally distributed noise with
two dispersions: one representing the variability of the star during
a night and one representing the night-to-night variability of the
star. The former was estimated by taking the mean of each night's
standard deviation, and the latter by taking the standard deviation
of nightly means. With this approach, the test light curves can
vary on timescales that are long compared to our sampling interval,
allowing them to mimic the random slow variability of stellar origin
that could produce spurious periodic behavior over our limited observing
window. The maximum power of the 100 periodograms of the test light
curves was adopted as the level of 1\% FAP, and used as the initial
threshold for detecting significant photometric variability. For all
stars that met the FAP criterion we examined (by eye) the periodogram
and raw and phased light curves. We report stellar rotation periods
for 441 stars in our database (see Table 1 in Appendix B). 
  
\begin{figure}[ht!]
\epsscale{1.0}
\plotone{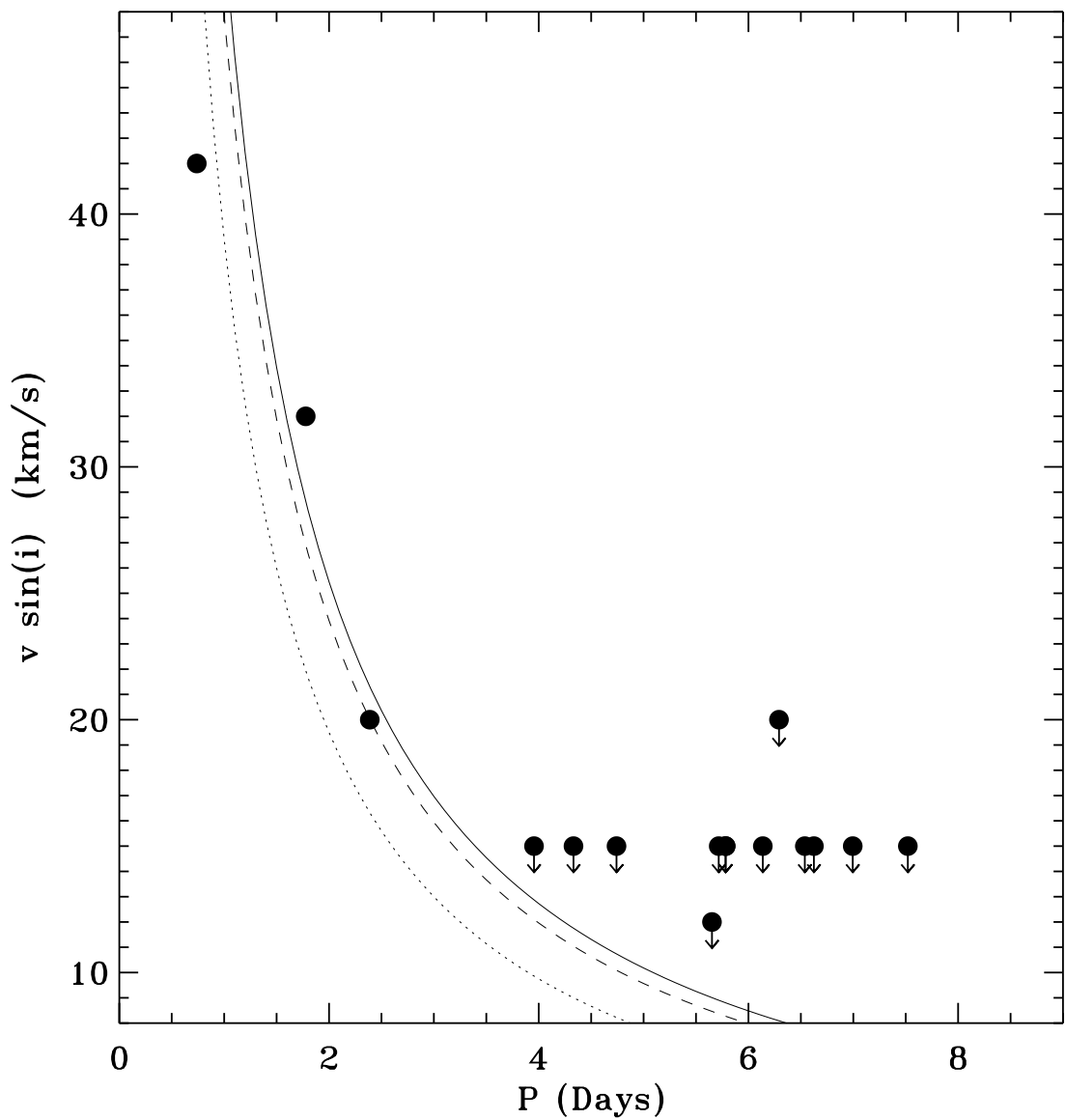}
\caption{Projected rotation velocities \citep{nds01} plotted
against the measured rotation period for 16 stars in M35. All
stars have $P_{RV} \ge 60\%$ and none of the 16 stars are spectroscopic
binaries. For comparison, the solid, dashed, and dotted curves
indicate the relation between rotation period and the projected
rotational velocity for a solar-like star with a $90\degr$,
$70\degr$, and $50\degr$ inclination of the rotational axis,
respectively. The rotation periods and the projected rotation
velocities are consistent for all 16 stars.
\label{vsini}}
\end{figure}

We do not have multiple seasons of observations or observations
in multiple pass-bands at our disposal by which to confirm rotation
periods of individual stars. However, the reliability of the
derived periods is supported by an observed correlation between
photometric period and rotational line broadening within a subset
of 16 single cluster members. Figure~\ref{vsini} shows the projected rotation
velocities measured by \citet{nds01} for 16 stars for which we
have determined rotation periods. The shortest period stars
($P_{r} \la 2.5$ days) show increasing $v \sin(i)$ with decreasing
rotation period. For rotation periods of $\sim$ 4 days or longer
the upper limits on the projected rotation velocities are consistent
with slower rotation. For comparison, the solid, dashed, and dotted
curves in Figure~\ref{vsini} indicate the relation between rotation
period and the projected rotational velocity for a solar-like star
with a $90\degr$, $70\degr$, and $50\degr$ inclination of the rotational
axis, respectively. Thus, for all 16 stars, the projected rotation
velocities are consistent with the measured rotation periods.

%\clearpage
%\clearpage

% -----------------------------------------------------------------------------

\subsection{The Spectroscopic Survey}			\label{spec}

M35 has been included in the WIYN Open Cluster Study
(WOCS; \citet{mathieu00}) since 1997. As part of WOCS
more than 6000 spectra has been obtained of approximately
1500 solar-type stars within a 1-degree field centered on
M35. The selection of survey target stars was based on
photometric (\citet{deliyannis08}, see Section~\ref{phot})
and proper-motion \citep{ms86,cudworth71} membership data. 
Stars on or less than $\sim 1\fm0$ above the cluster main sequence
were selected, with brightness and color ranges corresponding to
a range in mass from $\sim0.7~M_{\odot}$ to $\sim1.4~M_{\odot}$.
All spectroscopic data were obtained using the WIYN 3.5m
telescope equipped with a multi-object fiber optic positioner
(Hydra) feeding a bench mounted spectrograph. Observations were
done at central wavelengths of 5130\AA\ or 6385\AA\ with a wavelength
range of $\sim$200\AA\, providing many narrow absorption
lines. Radial velocities with a precision of $\la 0.5~km~s^{-1}$
\citep{gmh+08,mbd+01} were derived from the spectra via
cross-correlation with a high $S/N$ sky spectrum. From this
extensive radial-velocity survey we have 1) calculated the cluster
membership probability; 2) detected the cluster binary stars; and 3)
determined the orbital parameters for the closest binaries.

Of the 441 stars with rotation periods presented in this
study, 259 have one or more radial-velocity measurements
(the remainder being below the faint limit of the spectroscopic
survey or photometric non-members). The radial-velocity
cluster membership probability of each star is calculated
using the formalism by \citet{vkp58}. The mean or center-of-mass
radial velocity of a single or binary star was used when calculating
the membership probability. We have adopted 50\% as the threshold
for assigning radial-velocity and proper-motion cluster membership.
Of the 259 rotators with one or more radial-velocity measurement,
203 are radial-velocity members of M35 and 20 of those 203 stars
are also proper-motion members. More detailed descriptions of the
radial-velocity survey and membership determination can be found
in \citet{mm05}, \citet{mms06}, and \citet{bmm08}.

% -----------------------------------------------------------------------------

\subsection{The M35 Color-Magnitude Diagram and Photometric Membership}
\label{phot}

\begin{figure}[ht!]
\epsscale{1.0}
\plotone{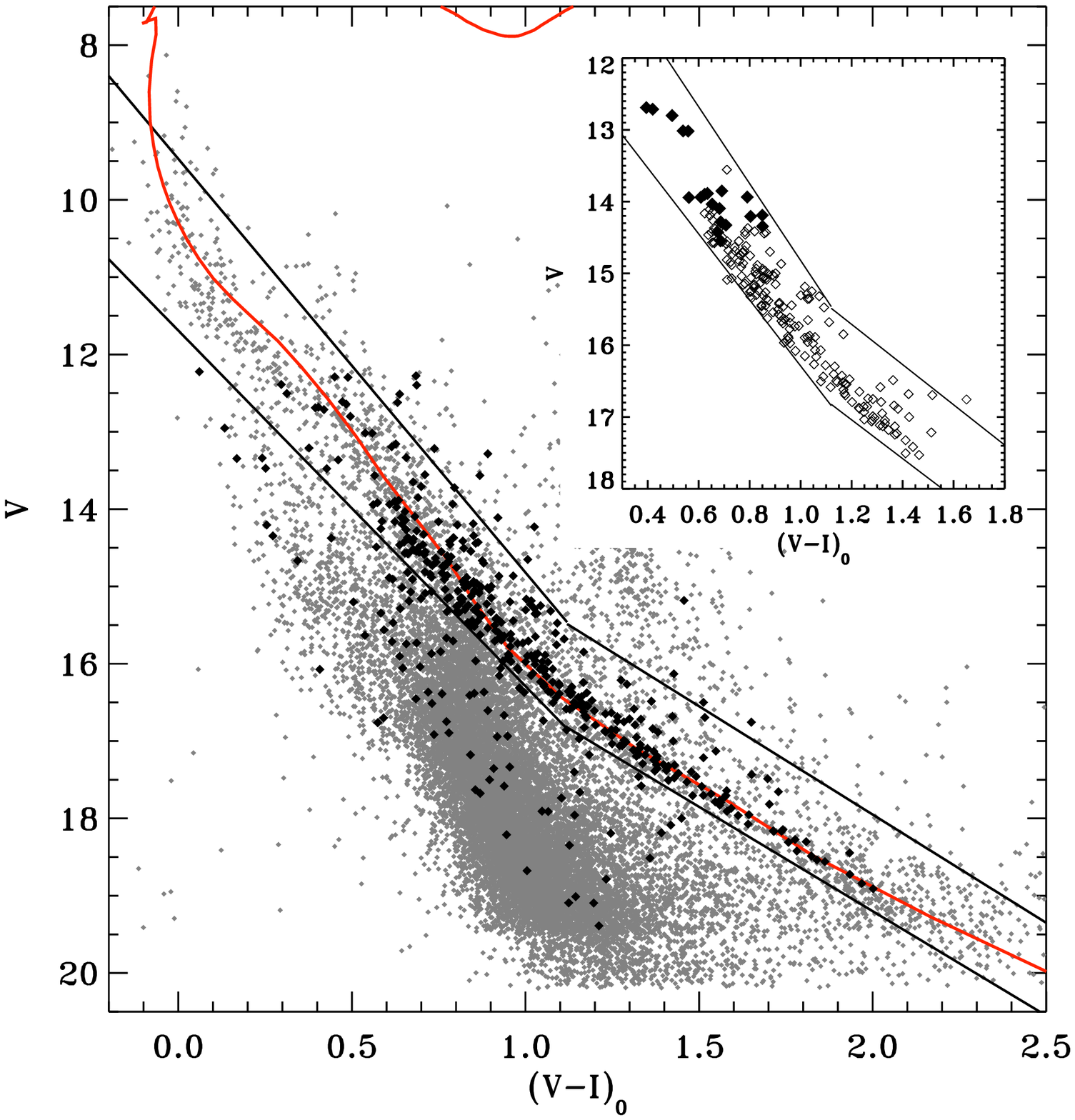}
\caption{The M35 $(V-I)_0$ vs. $V$ color-magnitude diagram.
Photometry was provided by \citet{deliyannis08}. The 441 stars
with rotation periods are highlighted in black. Stars located between
the solid lines are considered photometric members of M35. Note that
the faint limits for proper-motion and radial-velocity surveys are
$V \simeq 14.5$ and $V \simeq 17.5$, respectively. The insert shows
the location of stars that are radial-velocity members (open symbols),
and radial-velocity and proper-motion members (filled symbols). These
kinematic members of M35 were used to define the boundaries for photometric
membership. The isochrone shown represents a 150 Myr Yale model
\citep{ykd03}.
\label{cmd}}
\end{figure}

Figure~\ref{cmd} shows the (V-I) vs. V color-magnitude diagram (CMD)
for M35. The photometry was kindly provided by \citet{deliyannis08}
who obtained UBVRI data in a $23\arcmin \times 23\arcmin$ central
field and BVRI data in a $2 \times 2$ mosaic for a total of $\sim40\arcmin
\times 40\arcmin$ using the WIYN 0.9m telescope. In the CMD the 441
stars for which we have measured rotation periods are highlighted in
black. The solid lines enclosing stars within or above the cluster
sequence (allowing for inclusion of equal-mass binaries) show our criteria
for photometric membership. The insert in Figure~\ref{cmd} shows the
location in the M35 CMD of only radial-velocity members (open symbols),
and radial-velocity and proper-motion members (filled symbols),
the location of which was used to define the criteria for photometric
membership. There are 23 photometric members with 3 or more radial-velocity
measurements and a radial-velocity membership probability of less
than 50\%. Those 23 stars were removed from the list of cluster
members. The final number of stars with measured rotation periods
selected as radial-velocity and/or photometric members of M35 is 310.

\placefigure{cmd}

%\clearpage
%\clearpage

% *****************************************************************************

\section{THE ROTATION-PERIOD DISTRIBUTION}		\label{distr}

\begin{figure}[ht!]
\epsscale{1.0}
\plottwo{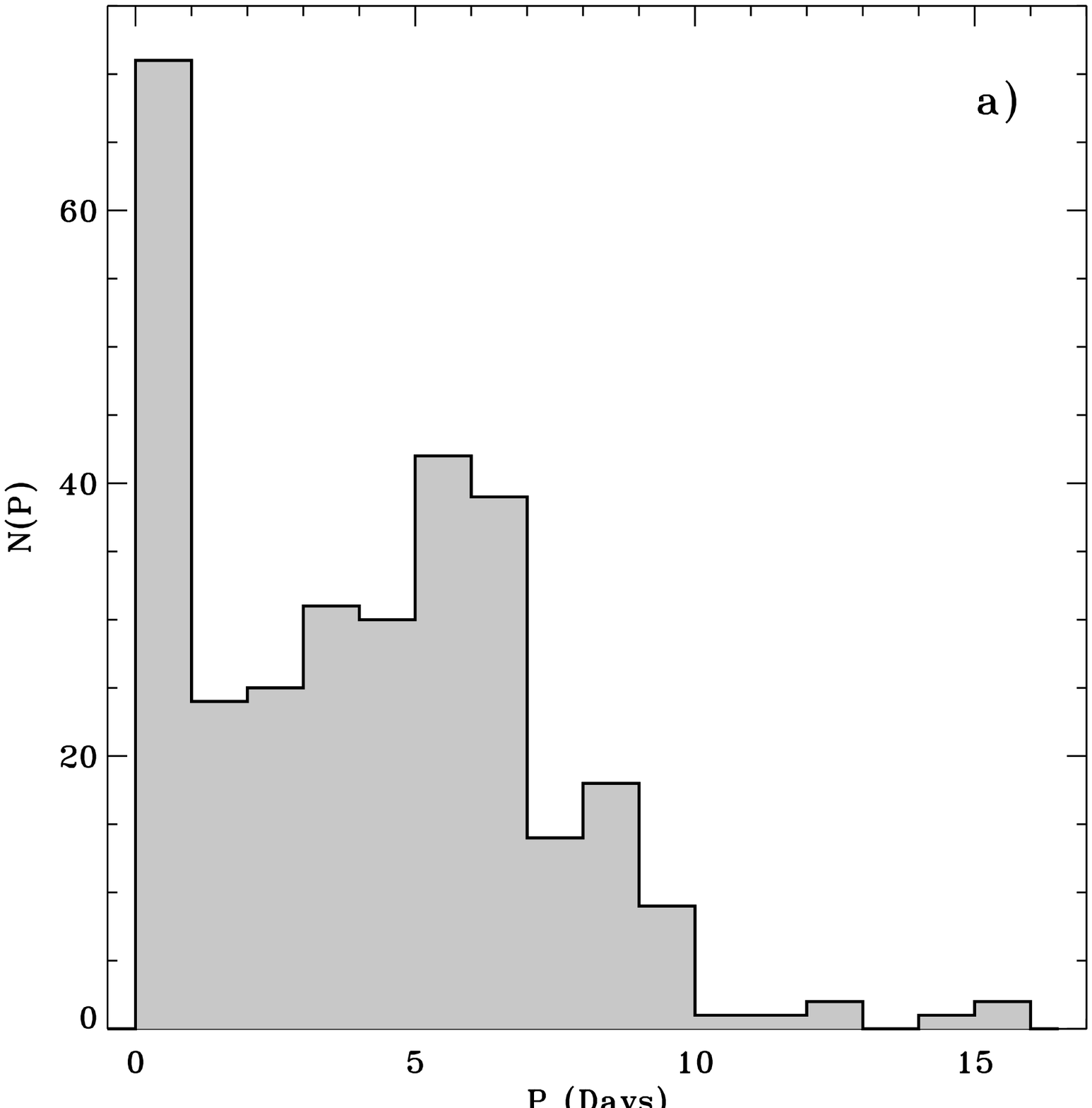}{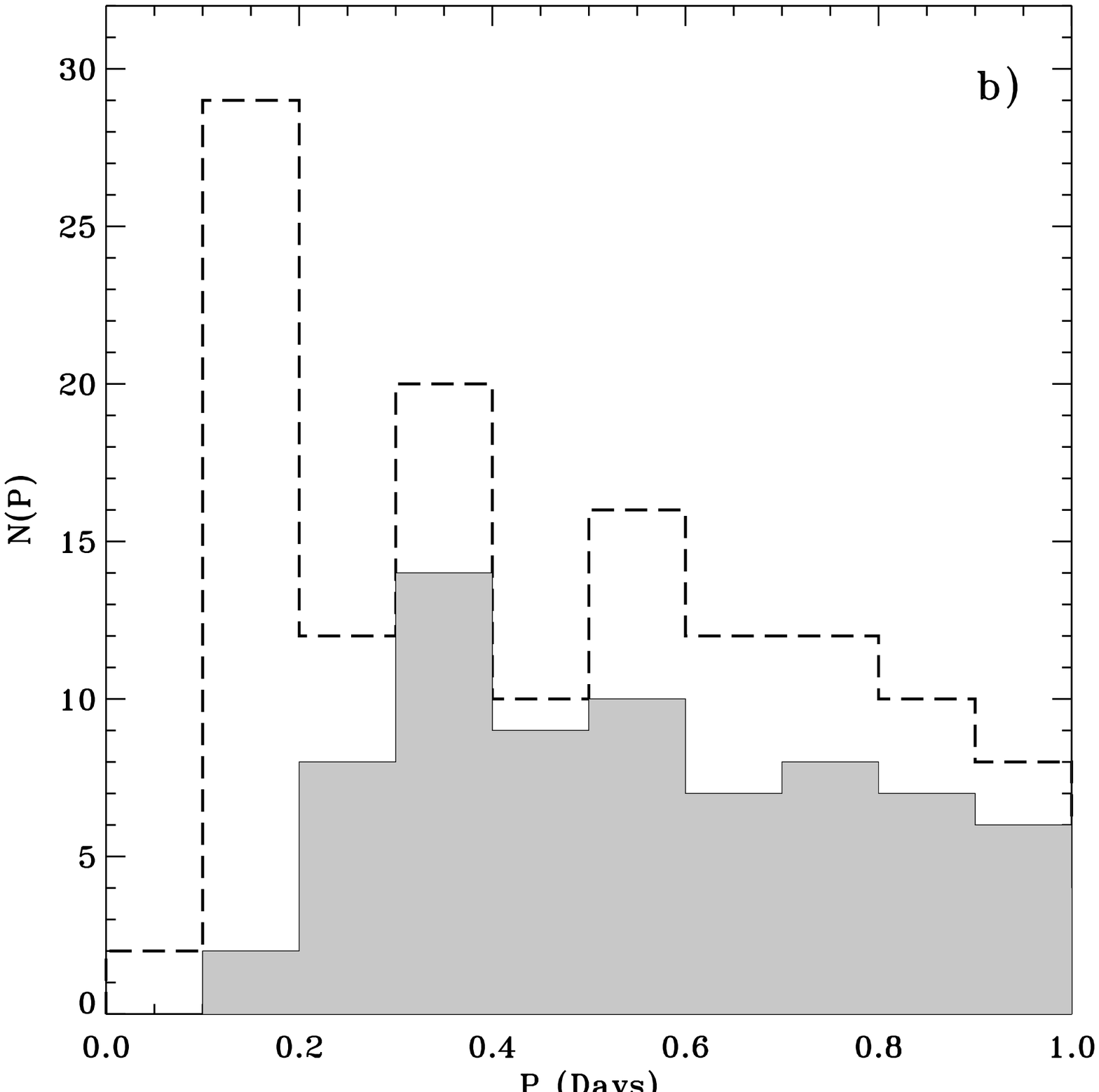}
\caption{{\bf a)} The distribution of rotation periods for the sample
of 310 cluster members with masses from $\sim0.6-1.4~M_{\odot}$ (spectral
type late K to mid F). The distribution show a large dispersion from
$\sim$0.1 days to $\sim$15 days, and a distinct peak at $\le 1$ day
and a shallower and broader peak centered at $\sim$6 days.
{\bf b)} The distribution of rotation periods shortward of 1 day binned
in 0.1 day bins. The dashed line histogram represents member as well as
non-member stars with measured rotation periods in our sample. The
grey histogram represents only members of M35.
\label{phist}}
\end{figure}

The 310 members of M35 with determined rotation periods correspond to
$\sim$12\% of the photometric cluster population within the brightness
and areal limits of our photometric survey. Figure~\ref{phist}a shows
the distribution of rotation periods, which spans more than 2 orders of
magnitude from $\sim$0.1 days to $\sim$15 days. The distribution peaks
shortward of 1 day and has a broader and shallower peak centered at about
6 days.

\placefigure{phist}

%\clearpage
%\clearpage

Figure~\ref{phist}b displays with an increased resolution of 0.1 day
the distribution of rotation periods shortward of 1 day. The dashed
and grey histograms, respectively, represent all stars and all cluster
members with detected rotation periods. The distribution
shows that we are capable of detecting rotation periods down to the
pseudo-Nyquist period-limit of about 2 hours ($\sim 0.08$ day) resulting
from our typical sampling cadence of about $1~hr^{-1}$ in December 2002.
The distribution of rotation periods for cluster members falls off shortward
of 0.3-0.4 days. Two member stars have rotation periods between 0.1 days and
0.2 days, corresponding to surface rotational velocities of 50\% or more
of their breakup velocities ($v_{br} = \sqrt{(GM_{\star}/R_{\star})}$).
We argue based on this inspection of the short-period tail of the
distribution that the lower limit of 0.1 days for our period search was
set appropriately for the stars in M35.

The long-period ends of the period distributions for members and
non-members (Figure~\ref{nmd}, Appendix~\ref{nm}) show that the
long time-span of the queue-scheduled data enable us to detect rotation 
periods beyond the $\sim$10 days typically found to be the upper limit
in photometric surveys with durations similar to our short-term December
2002 observing run. We report the detection of 18 stars with rotation
periods longer than 10 days, 7 of which are members of M35. The longest
rotation period among members is 15.3 days, and among the field stars
rotation periods of up to $\sim$ 17 and 23 days have been measured.
In the M35 period distribution
we see a drop-off at $\sim$10 days. If the $\sim$12\% of the cluster's
late-type population with measured rotation periods is a representative
sample of the late-type stellar population in M35, then the $\sim$10 day
cutoff may represent a physical upper limit on the rotation-period
distribution at 150 Myr. However, it is also possible that we are not
capable of detecting the slowest rotators despite our long-baseline
photometric survey. Indeed, the modest number of rotation periods longer
than 10 days found in the much larger sample of field stars may reflect
that the frequency and size of spots on stars rotating slower than
$\sim$10 days is insufficient for detection with the photometric precision
of our data. Indeed, X-ray observations in Orion \citep{sab+04} indicate
that rotation-period studies of young stars may in general be biased
against very slow rotators because such stars likely do not generate
strong activity, and thus are not sufficiently spotted to allow detection of
photometric periods."
Measuring the rotation for such slowly rotating stars will likely
require either higher photometric precision or high resolution ($R \ga
50,000$) spectroscopic observations to measure projected rotation velocities.
The small sample of M35 member stars with periods above 10 days will
be discussed in Section~\ref{prediction}.

% -----------------------------------------------------------------------------

\subsection{Long-term stability of the number, sizes, and configurations
of stellar spots and spot-groups}

We find that for almost all cluster and field stars with measured rotation
periods, the long-term queue-scheduled data, spanning $\sim$5 months in time,
phase up with and coincide with the short-term data (16 nights in December
2002) in the light curves. We tested further the agreement between the
short-term and long-term photometric variability, by measuring the rotation
period separately from the short- and long-term data for 20 randomly chosen
stars. For all but one star we found a difference between the two rotation
periods that was less than 1\% of the period measured from the short-term
data alone (in most cases the difference was less than 0.1\%). We examined
the 20 light curves with all data phased to the period derived from only
the short-term data. For all but one star, the long-term data produced light
curves of the same shape and phase as the light curves based only on the
short-term data. Even for two light curves with clear signs of multiple
spots (spot-groups), the short- and long-term data coincided very well.
Because it is unlikely that the disappearance and recurrences of
spots will result in a light curve with the same shape, amplitude, and
phase, the agreement suggests stability of individual spots and/or
spot-groups over the $\sim$5 month time-span of our photometric observations.

We find that the stability of the sizes and configurations of spots on
young stars have recently been studied for e.g. the solar analog PMS star
V410 Tau \citep{sfc+03} and for stars in the PMS cluster IC348
\citep{nhr+06}. For V410 Tau data has been collected for over two decades,
showing changes in the shape of the light curve over the last decade.
The authors suggest that the observed changes reflect variations of the
structure of the active regions over timescales of years. However, stability
in the rotation period and the recurrence of the light curve minimum,
suggest stability of the largest spots over years, and either a lack
of latitudinal differential rotation in V410 Tau, or confinement of its
spots to a narrow range of latitudes.  Similarly, \citet{nhr+06} finds
a remarkable stability over 7 years in the rotations periods for stars
in IC348, suggesting again that these PMS stars do not have significant
differential rotation, or that their spots are constrained to a narrow
range of stellar latitudes. However, contrary to what is observed over
5 months in M35, all periodic stars in IC348, as well as V410 Tau, do
show changes in the light curve shape and amplitude from year to year.

% *****************************************************************************

\section{THE M35 COLOR-PERIOD DIAGRAM - THE DEPENDENCE OF STELLAR
ROTATION ON MASS} \label{cpd}

\begin{figure}[ht!]
\epsscale{1.0}
\plotone{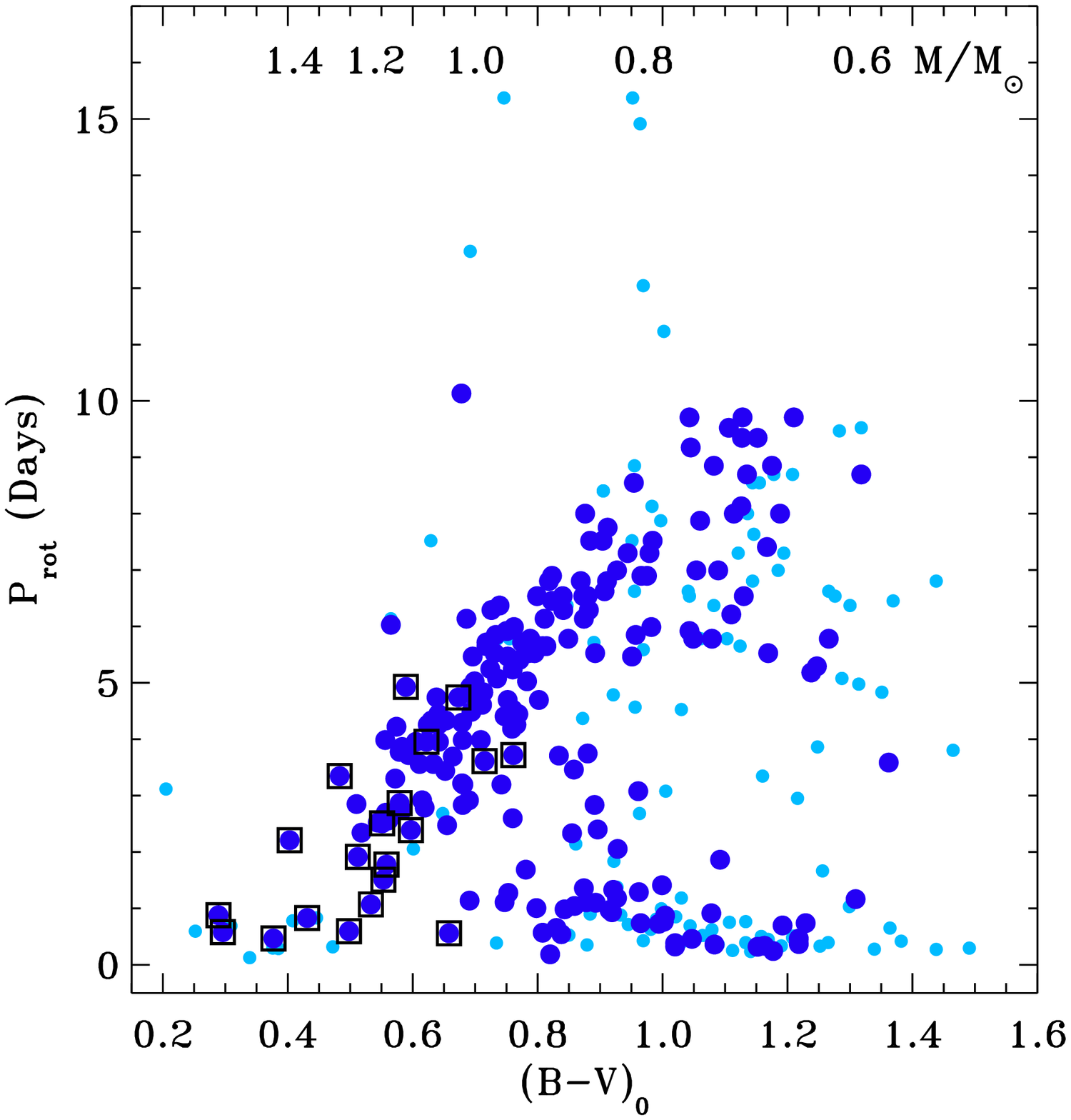}
\caption{The distribution of stellar rotation periods with (B-V) color
index for 310 members of M35. Dark blue symbols represent stars that are
both photometric and radial-velocity members of M35. Light blue symbols
are used for stars that are photometric members only. Proper-motion members
are marked with additional squares. The upper x-axis gives a stellar mass
estimate corresponding to the color on the lower axis. Masses are derived
using a 150 Myr Yale isochrone.
\label{pbv}}
\end{figure}

In Figure~\ref{pbv} we display the rotation periods for the 310 members
plotted against their $B-V$ color indices, or equivalently their
masses. The color indices derive from the deep multi-band photometry
by \citet[][Section~\ref{phot}]{deliyannis08} and the corresponding
stellar mass estimates (upper x-axis) from a fit of a 150\,Myr Yale
stellar evolutionary model \citep{ykd03} to the M35 color-magnitude diagram.
Dark blue symbols represent stars that are both photometric and
radial-velocity members of M35. Light blue symbols are used for stars
that are photometric members only. Proper-motion members are marked
with additional squares.

\placefigure{pbv}

%\clearpage
%\clearpage

The M35 color-period diagram shows striking structure. The coeval stars
fall along two well-defined sequences apparently representing two different
rotational states. One sequence displays clear dependence between
period and color, starting at the blue end at $(B-V)_0 \simeq 0.5$
($M_{\star} \simeq 1.2~M_{\odot}$) and $P_{rot} \simeq 2$ days and forming
a rich diagonal band of stars whose periods are increasing with increasing
color index (decreasing mass). This sequence terminates at about $(B-V)_0
\simeq 1.2$ ($M_{\star} \simeq 0.65~M_{\odot}$) and $P_{rot} \simeq 10$ days.
The second sequence consists of rapidly rotating ($P_{rot} \la 1$ day) stars
and extends from $(B-V)_0 \simeq 0.7-0.8$ ($M_{\star} \simeq
0.9-1.0~M_{\odot}$) to $(B-V)_0 \simeq 1.5$ ($M_{\star} \la 0.5~M_{\odot}$).
This well defined sequence of rapidly rotating stars shows a small but
steady decrease in rotation period with increasing color (decreasing mass).
Finally, a subset of stars are distributed in between the two sequences,
and 10 stars have rotation periods that are unusually long, placing them
above the diagonal sequence in Figure~\ref{pbv}.

The M35 color-period diagram gives a clear picture of preferred
stellar rotation periods as a function of color for 150\,Myr late-type
dwarfs. With the added dimension of color, the diagram
take us beyond the one-dimensional period distribution and shows
which stars are responsible for the structure observed in
Figure~\ref{phist}. The short-period peak at $\la$1 day is due primarily
to the rapidly rotating late G- and K-dwarfs ($M \la 0.9\,M_{\odot}$),
and the sharpness of this peak is the result of little dependence
of rotation on color within this group. The more slowly rotating
mid to late G- and K-dwarfs give rise to the broader peak at $\sim$6
days, while the early to mid G-dwarfs and some cooler stars fill in
the distribution between the peaks.

The two sequences of stars in the color-period diagram
represent the most likely/stable rotation period(s) for a given stellar
mass at the age of M35. Under the assumption that rotation periods
increase with time for all stars, the more sparsely populated area
between the two sequences must then represent a phase of rotational
evolution of shorter duration. We will discuss the different loci
in the color-period diagram in more detail in Section~\ref{discuss}.

% *****************************************************************************

\section{ANGULAR MOMENTUM EVOLUTION AND THE COLOR-PERIOD DIAGRAM}
\label{discuss} 

Sequences similar to those observed in the M35 color-period diagram
were noted by \citet[][hereinafter B03]{barnes03a} from careful examination
of compilations of rotation-period data from photometric monitoring
campaigns on open clusters and field stars. B03 named the diagonal
sequence of stars on which rotation periods increase with color the
{\it interface} sequence (or I sequence), and the sequence of rapidly
rotating stars the {\it convective} sequence (or C sequence). In what
follows we will adopt these names for the two sequences in the M35
color-period diagram. 

B03 argues that the rapidly rotating stars on the C sequence have
radiative cores and convective envelopes that are decoupled. For
these stars he suggests that the evolution of their surface rotation
rates is governed primarily by the moments of inertia of the convective
envelope and by inefficient wind-driven loss of angular momentum linked
to small-scale convective magnetic fields. For stars on the I sequence,
large-scale (sun-like) magnetic fields provided by an interface dynamo
couple the core and envelope, and the rotational evolution of the I
sequence stars is thus primarily governed by the moments of inertia
of the entire star and more efficient angular momentum loss (i.e., a
\citet{skumanich72} style spin-down).
Accordingly, B03 suggests that a late-type star,
in which the core and envelope are decoupled as it settles on the ZAMS,
will begin its main-sequence life on the C sequence and evolve onto
the I sequence when rotational shear between the stellar core and
envelope establish a large-scale dynamo field that couples the two zones
and provide efficient magnetic wind loss. Higher mass stars have thinner
convective envelopes with smaller moments of inertia than low mass stars
and thus leave the C sequence sooner. Stars that are either fully radiative
or fully convective will remain as rapid rotators.

Color-period diagrams for coeval populations of different ages
allow us to examine the rotational properties of late-type coeval stars
as a function of their mass and age, and may bring us closer to understanding
the physical mechanisms (internal and/or external) regulating their
rotational evolution. The M35 color-period diagram, rich in stars and
cleaned for spectroscopic and photometric non-members, reveals the
morphology described by B03 more clearly than any published stellar
populations. We therefore begin with a discussion of the M35
result in the context of the framework developed by B03.

% -----------------------------------------------------------------------------

\subsection{Timescales for migration from the C to the I sequence}
\label{timescales}

\begin{figure}[ht!]
\epsscale{1.0}
\plotone{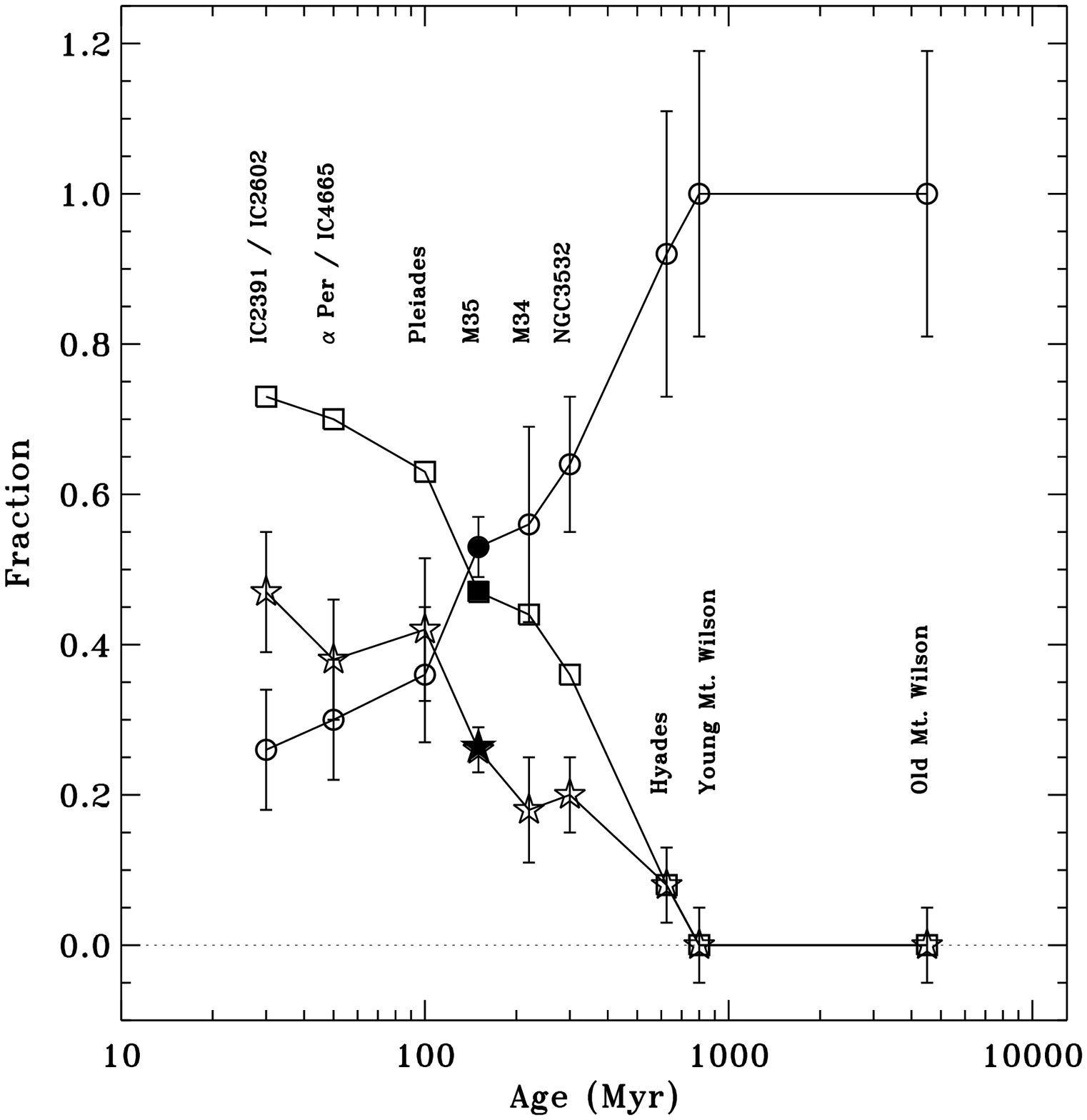}
\caption{Figure 3 from B03 with M35 added. The figure shows
the fractions of stars with $0.5 \le (B-V)_0 \le 1.5$ on the I sequence
(circles) and the C sequence (stars) for clusters of different ages.
The squares represent the relative fraction of the sum of C sequence
and gap stars. The filled symbols show the relative fractions for M35.
We follow B03 in estimating the uncertainties in the
fractions by the square root of the number of stars.
\label{fractions}}
\end{figure}

In Figure~\ref{fractions} we add M35 to Fig. 3 in B03 which shows the
relative fractions of stars with $0.5 \le (B-V)_0 \le 1.5$ on the I and
C sequences for stellar populations of distinct ages. With rotation periods
measured over more than two orders of magnitude for confirmed
spectroscopic and photometric cluster members, M35 adds the statistically
most secure datapoints to this figure. M35 fit well with the evolutionary
trends of increasing relative fractions of C sequence (and gap) stars
and decreasing fractions of I sequence stars for younger cluster populations.
The almost linear trends in Figure~\ref{fractions} suggest that the
decrease and increase in the number of stars on the C and I sequences,
respectively, are approximately exponential with time. Under the presumption
of an exponential time dependence and that all stars start on the C sequence
at the ZAMS,
we can estimate the characteristic timescale for the rotational evolution
of stars off the C sequence and onto the I sequence, by counting stars on
both sequences and in the gap in the M35 color-period diagram. Such
timescales may offer valuable constraints on the rates of internal and
external angular momentum transport and on the evolution rates of stellar
dynamos in late-type stars of different masses.

\placefigure{fractions}

%\clearpage
%\clearpage

When counting the number of stars on the I and C sequences and in the gap,
we use the following criteria. Stars located in the color-period diagram
between the lines represented by $P_{rot} = 10 (B-V)_0 - 2.5 \pm 2.0$ and
with periods above 1.5 days were counted as I sequence stars (see dotted
lines in Figure~\ref{pbv_iso2}). Stars redder than $(B-V)_0 = 0.6$ and with
periods between 0 and 1.5 days were counted as C sequence stars. Stars
located below $P_{rot} = 10 (B-V)_0 - 4.5$ and with periods above 1.5 days
were counted as gap-stars. These selection criteria are subjective and
although the sequences are well defined, the I sequence becomes broader
redward of $(B-V)_0 \simeq 1.0$ making the distinction between I sequence
and gap stars more difficult. However, due to the large number of rotation
periods in M35, the small number of stars that might be moved from the gap
to the I sequence or vice versa by using slightly different criteria will
not influence the relative fractions and thus the timescales in any
significant way.

The number of C sequence stars ($N_c$) at a time $t$ can then be expressed
by:

\begin{equation}
N_c = N_{c_{0}} e^{-t/\tau_c}
\end{equation}

\noindent where $N_{c_{0}}$ and $\tau_c$ are, respectively, the total number
of stars on the I and C sequences and in the gap, and the characteristic
exponential timescale. We use the B-V color index to divide the stars into
G-dwarfs ($0.6 < B-V < 0.8$) and K-dwarfs ($0.8 < B-V < 1.3$) We count all
stars within each color-interval in the color-period diagram as $N_{c_{0}}$,
all C sequence stars within each color-interval as $N_c$, and adopt
150\,Myr as the age of M35. We derive from equation [1] $\tau_c^G = 60$ Myr
and $\tau_c^K = 140$ Myr as the characteristic exponential timescales for
transition between the C and the I sequence for G and K dwarfs, respectively.

We can qualitatively verify these time scales by a comparison between the
M35 color-period diagram and those of the younger Pleiades cluster and
the older cluster NGC3532 presented in B03. In M35 only 7 G dwarfs ($0.6
\la (B-V)_0 \la 0.8$) are found on the C sequence and in the gap, while the
G dwarf I sequence is well defined and rich. In contrast, In contrast, the
M35 C sequence and gap are rich in K dwarfs ($0.8 \la (B-V)_0 \la 1.3$),
whereas the K dwarf I sequence is less densely populated and less well defined.
The lack of G dwarfs on the C sequence seen in M35 is already apparent at
100 Myr in the Pleiades color-period diagram, indicating that the
characteristic timescale for G dwarfs to evolve off the C sequence
and onto the I sequence is less than $\sim$100-150 Myr. The rich population
of early and mid K dwarfs on the M35 C sequence have evolved off the C
sequence and onto a well defined I sequence by the age of NGC\,3532 (300 Myr).
The NGC\,3532 C sequence and gap, however, are populated by late K dwarfs,
suggesting that early to mid K dwarfs evolve onto the I sequence on a
timescale between 150 and 300 Myr, or approximately twice the time
required for G dwarfs. Finally, by the age of the Hyades only 3 late K or early
M dwarfs have been found off the I sequence, or in the gap (see Fig. 1
in B03 or Figure~\ref{m35_hyades_pbv} below), suggesting that such stars
evolve off the C sequence and possibly onto the I sequence on a timescale
of $\sim$600 Myr, or approximately twice the time required for the early
to mid K dwarfs.

There is thus good agreement between the exponential timescales derived
from the M35 color-period diagram alone and the estimated timescales based
on a comparison of color-period diagrams of different ages. We note that
although the relative fractions displayed in Figure~\ref{fractions} represent
all rotators with $0.5 \le (B-V)_0 \le 1.5$, a least-squares fit of an
exponential function to the C sequence fractions in Figure~\ref{fractions},
gives a timescale of 106\,Myr for a decrease in the number of C sequence
stars by a factor of $e$. 

% -----------------------------------------------------------------------------

\subsection{Testing the Skumanich $\sqrt{t}$ spin-down rate between
M35 and Hyades}	\label{skumanich} 

\begin{figure}[ht!]
\epsscale{1.0}
\plotone{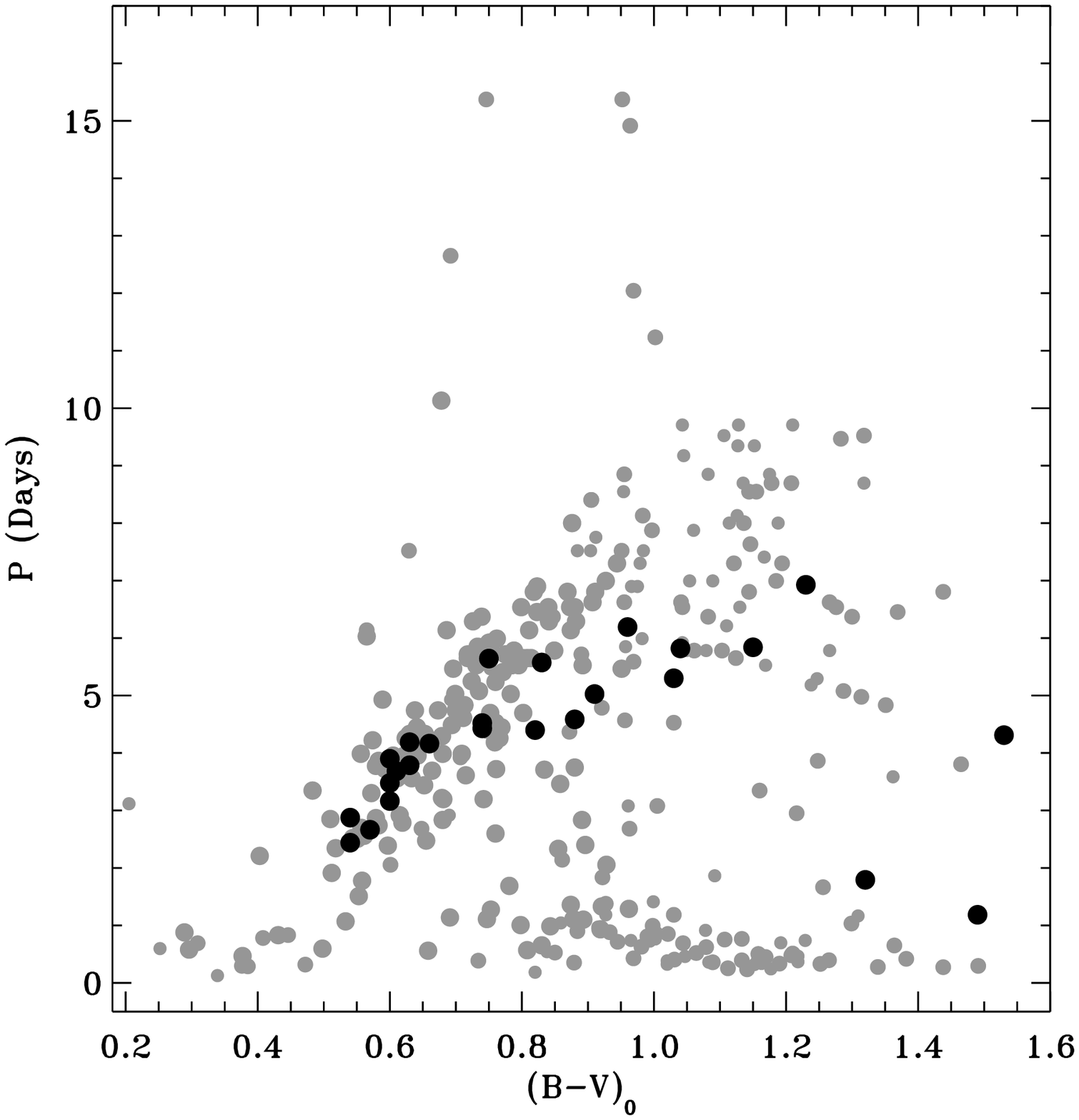}
\caption{The M35 color-period diagram (grey symbols) with 25 Hyades stars
overplotted \citep[black;][]{rtl+87,psd+95}. All but the 3 reddest Hyades
stars fall on a sequence similar to the M35 I sequence. All Hyades rotation
periods were spun-up by a factor $\sqrt{625/150} \simeq 2$ in accordance
with the Skumanich $\sqrt{t}$ time-dependence on stellar rotation evolution,
assuming ages of 625\,Myr and 150\,Myr for the Hyades and M35, respectively.
\label{m35_hyades_pbv}}
\end{figure}

The color-period diagram for the Hyades contains 25 stars
\citep{rtl+87,psd+95}, 22 of which form an I sequence of G and K dwarfs.
Despite the smaller number of stars, the blue part of the Hyades I sequence,
populated by G dwarfs, is well defined. By comparing the rotation periods
for the I sequence G dwarfs in M35 to those of the I sequence G dwarfs in
the $\sim$4 times older Hyades, we can directly test the \citet{skumanich72}
$\sqrt{t}$ time-dependence on the rotation-period evolution for stars in
this mass-range. We follow B03
in assuming separate mass and time dependencies for stellar rotation,
and that the same mass dependence can be applied to different stellar
populations. Adopting an age of 625\,Myr for the Hyades \citep{pbl+98}
and of 150\,Myr for M35, we decrease the Hyades rotation periods
by $\sqrt{625/150} \simeq 2$. We show in Fig.~\ref{m35_hyades_pbv}
the color-period diagram with the 310 M35 members (all grey symbols) and
with the locations of the 25 Hyades stars overplotted (black symbols).
The spun-up Hyades I sequence G dwarfs coincide nicely with the M35
I sequence G dwarfs, in support of the Skumanich $\sqrt{t}$ time-dependence
for such stars.
Curiously, the Hyades K-dwarfs, also spun-up according to the Skumanich
law, have rotation periods systematically shorter than the M35 K dwarfs.
At face value, this suggests that the time-dependence for spin-down of
K dwarfs is different and slower than for G dwarfs between 150\,Myr and
625\,Myr.

\placefigure{m35_hyades_pbv}

%\clearpage
%\clearpage

% -----------------------------------------------------------------------------

%\subsection{The gyro-age of M35 - fitting rotational isochrones to
%the M35 color-period diagram}	\label{isochrones}

\subsection{The gyro-age of M35}	\label{isochrones}

Arguing that the rotation of stars on the C and I sequences follow
separate mass (M) and age (t) dependencies ($P(t,M) = g(t) \times f(M)$),
B03 introduced heuristic functional forms to represent these separate
dependencies of the I and C sequences. \citet[][hereinafter B07]{barnes07}
presents a modified functional form for the I sequence. These
functions define one-parameter families, with that parameter being the
age of the stellar population, and the resulting curves in the color-period
plane represent a set of rotational isochrones. We note that B03 use
$g(t) = \sqrt{t}$ (Skumanich), while B07 derive $g(t) = t^{0.52}$ by
requiring that a solar-like star spin down to solar rotation at solar age.

\citet{kawaler89} used his own calibrated angular momentum loss law
\citep{kawaler88} and the assumption of solid body rotation after
$\sim$100\,Myr \citep{pks+89}, to derive a relationship between stellar
age, rotation period, and color. Kawaler's age-period-color relation
is thus based on models calibrated to the sun, and the assumption that the
Skumanich relationship is valid for all masses. He express the stellar
mass, radius, and moment of inertia, in terms of observables such as
stellar color, via stellar models.

We note that our result in Section~\ref{skumanich}, suggest that the time
dependence of stellar rotation is not independent of stellar mass, as
assumed by both B03, B07, and \citet{kawaler89}.

\begin{figure}[ht!]
\epsscale{1.0}
\plotone{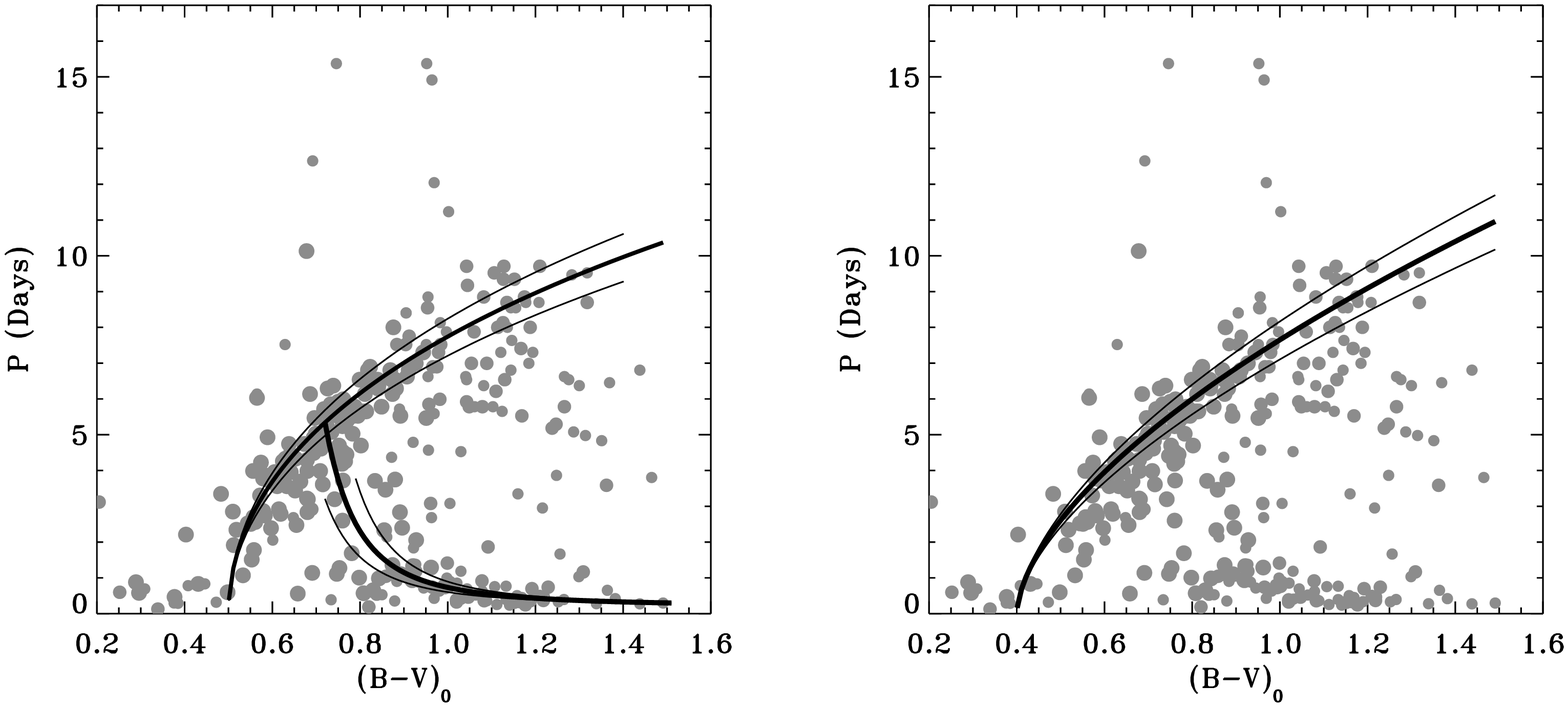}
\caption{The M35 color-period diagram with the 150\,Myr rotational isochrones
from B03 (left) and B07 (right) overplotted as a thick black solid curves.
To illustrate the sensitivity to age we show the 130\,Myr and the 170\,Myr
isochrones as thinner curves flanking the 150\,Myr isochrones.
\label{pbv_iso}}
\end{figure}

The well-defined sequences in the M35 color-period diagram make possible
a test of the period-color relations proposed by B03, B07, and
\citet{kawaler89}. We show in Figure~\ref{pbv_iso} the M35 color-period
diagram with the B03 and B07 rotational isochrones for the independently
determined stellar-evolution age for M35 of 150 Myr
\citep{vss+02,deliyannis08}. The rotational isochrones match the M35 I
and C sequences well, suggesting that they can indeed provide a consistent
age estimate (gyro-age) for a cluster based on a well populated color-period
diagram cleaned for non-members. To illustrate the sensitivity to age,
rotational isochrones for 130 Myr and 170 Myr are also displayed in
Figure~\ref{pbv_iso}.

\placefigure{pbv_iso}

%\clearpage
%\clearpage

\begin{figure}[ht!]
\epsscale{1.0}
\plotone{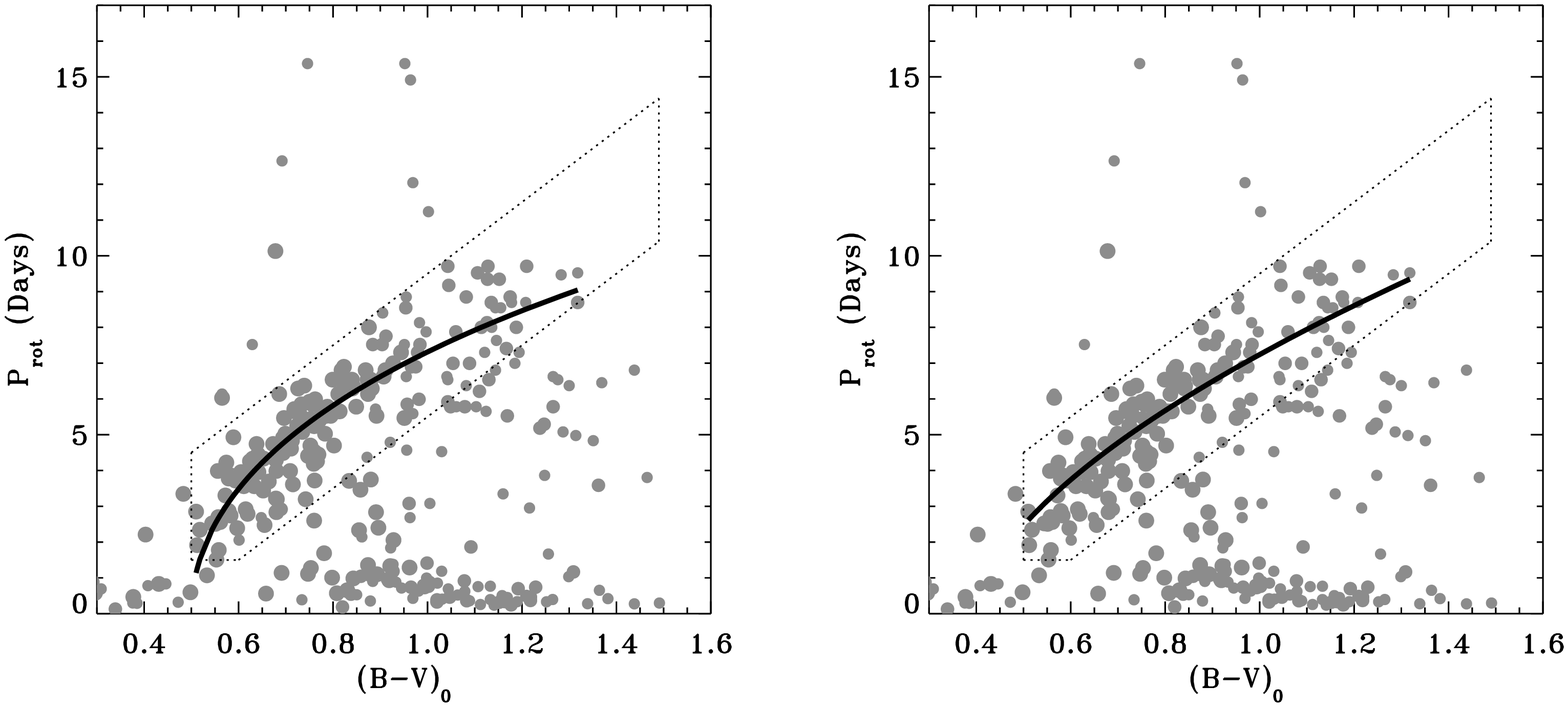}
\caption{The least squares fit of the B03 I sequence isochrone
(left; eq. [2]) and the B07 I sequence isochrone (right; eq. [3])
to the M35 I sequence with age ($t$) as a free parameter. The gyro-ages
corresponding to the fits are $133.9 \pm 3$Myr and $133.5 \pm 3$Myr,
respectively. The I sequence stars to which the isochrones were fitted
are enclosed by the dotted lines in both figures.
\label{pbv_iso2}}
\end{figure}

Assuming no prior knowledge about the age of M35 (thus letting age be a
free parameter), we perform a non-linear least squares fit to the I
sequence stars (enclosed by the dotted lines in Figure~\ref{pbv_iso2})
of the functional form of the rotational I sequence isochrones from
B03:

\begin{equation}
P(t,(B-V)) = \sqrt{t} \times \bigl(\sqrt{((B-V)-a)} - b((B-V)-a\bigr)
\end{equation}

\noindent with $a = 0.50$, and $b = 0.15$, and from B07:

\begin{equation}
P(t,(B-V)) = t^{0.52} \times \bigl(c((B-V)-d)^{f}\bigr)
\end{equation}

\noindent with $c = 0.77$, $d = 0.40$, and $f = 0.60$.

When fitting, the I sequence stars were weighted them according to their
cluster membership, with most weight given to confirmed radial-velocity
and proper motion members and least weight given to stars with only
photometric membership. The weights given to individual stars are listed
in Table~\ref{tab2} in Appendix B. Figure~\ref{pbv_iso2} shows the best
fits of both eq. [2] and eq. [3] to the M35 I sequence. The derived age
is 134\,Myr for both functional forms, each with a formal $1-\sigma$
uncertainty of $\sim$3\,Myr. The close agreement of the two ages likely
reflects the similar shape of the two isochrones over the color interval
from $\sim0.5-1.0$ where the M35 I sequence is most densely populated.
The small formal uncertainties reflect a well-defined I sequence rich
in stars. However, the gyro-ages may still be affected by systematic
errors in the stellar evolutionary ages for the young open clusters.

\placefigure{pbv_iso2}

%\clearpage
%\clearpage

\begin{figure}[ht!]
\epsscale{1.0}
\plottwo{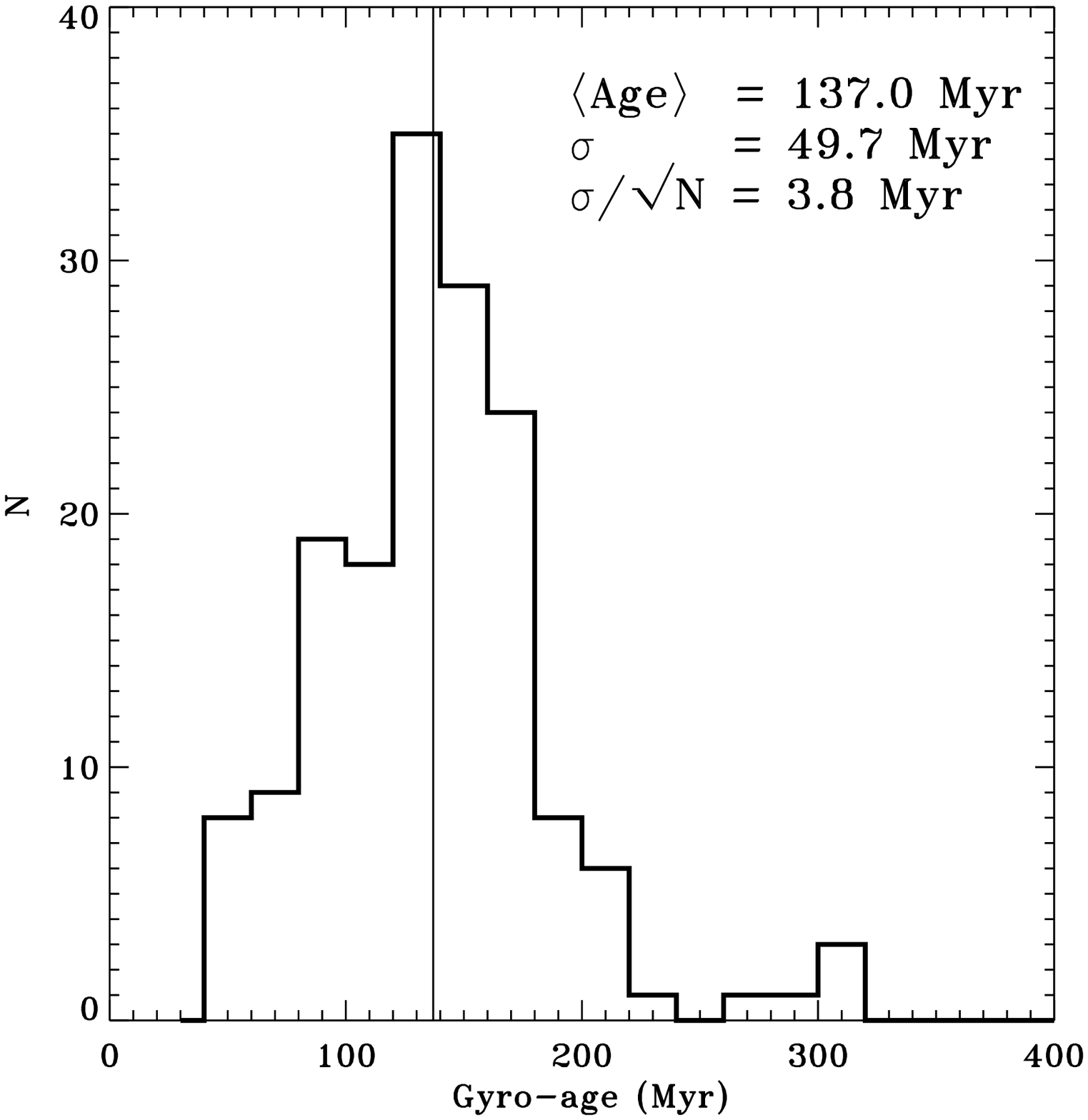}{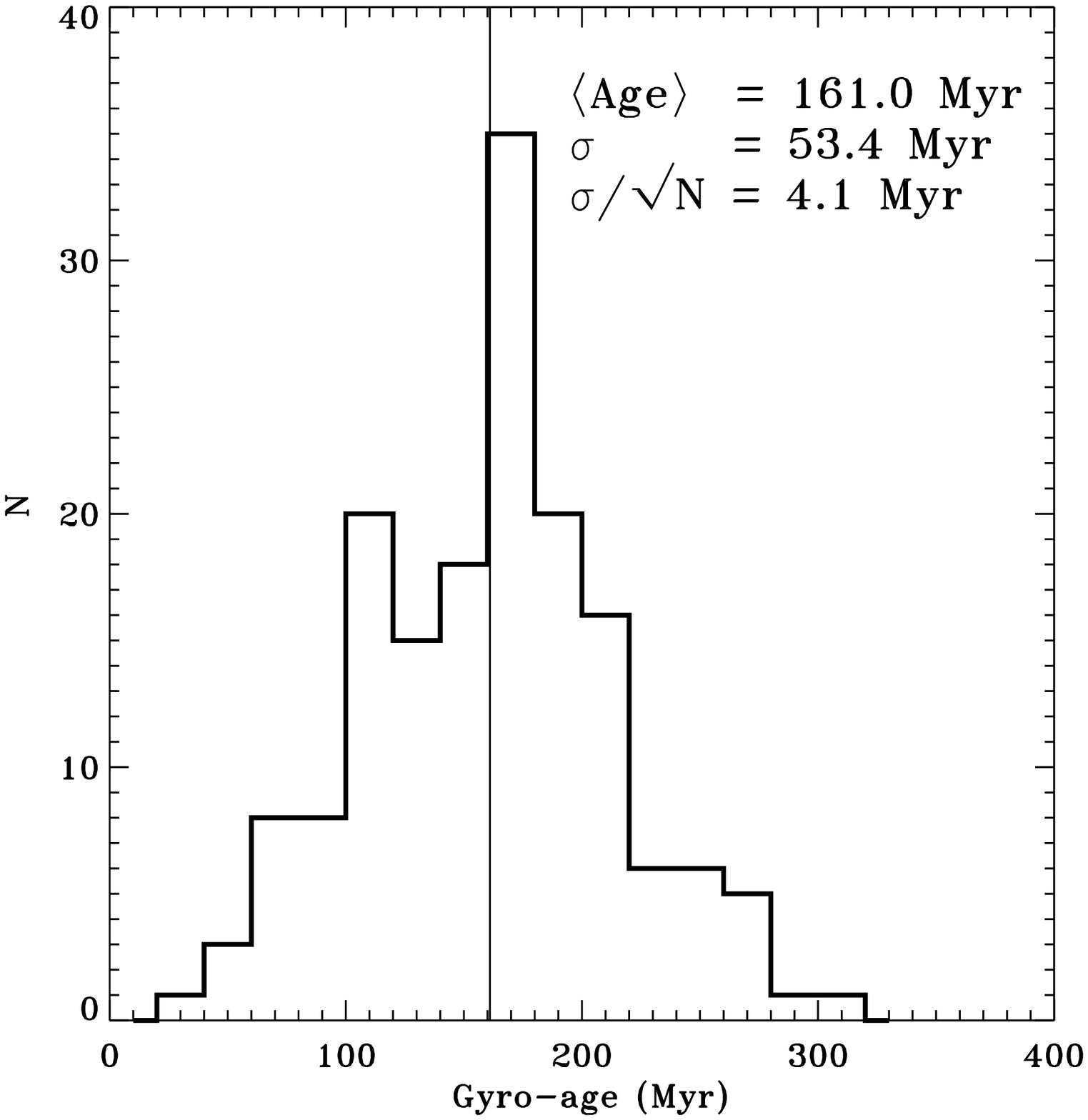}
\caption{The distribution of gyro-ages for M35 I sequence stars.
The panels show the distributions of M35 gyro-ages calculated using
the B03 (left panel) and B07 (right panel) age-rotation-color
relations. The distribution mean, standard deviation, and standard
error on the mean, are given in the upper right corner of each panel.
\label{age-distr}}
\end{figure}

Alternatively, we show in Figure~\ref{age-distr} the distribution
of gyro-ages of the M35 I sequence stars, calculated using the
age-period-color relations of B03 (left panel) and B07 (right panel).
The mean gyro-ages of 137\,Myr and 161\,Myr, are both close to the
150\,Myr derived for the cluster using the isochrone method (see
Section~\ref{intro}). Assuming that all I sequence stars are truely
coeval, the standard errors of 3.8\,Myr and 4.1\,Myr give uncertainties
on the calculated mean gyro-ages for M35 of 2.8\% and 2.5\%, close to
the formal uncertainty on the least squares fits. The close agreement
between the mean gyro-age (137\,Myr) and the gyro-age determined from
the least square fit (134\,Myr), as well as a smaller standard deviation
in the gyro-age distribution, suggest that the B03 I sequence isochrone
(eq. [2]) provide a better match to the color dependence of stellar
rotation on the I sequence than the B07 isochrone (eq. [3]).

\placefigure{age-distr}
 
%\clearpage
%\clearpage

\begin{figure}[ht!]
\epsscale{1.0}
\plotone{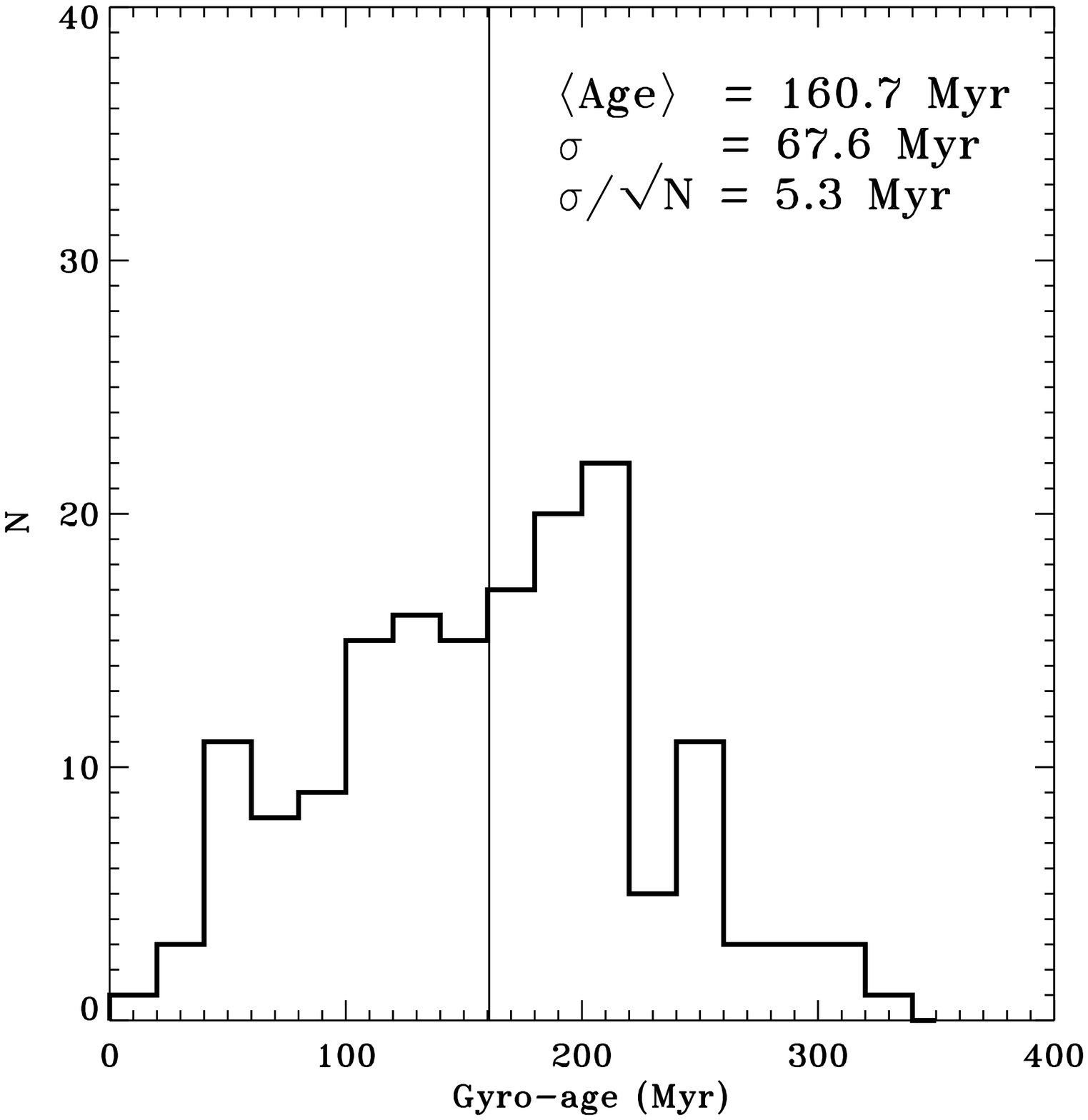}
\caption{The distribution of gyro-ages for M35 I sequence stars
calculated using the \citet{kawaler89} age-rotation-color relation.
The distribution mean, standard deviation, and standard error on the
mean, are given in the upper right corner of each panel.
\label{age-dist_kawalerr}}
\end{figure}

In Figure~\ref{age-dist_kawalerr} we show the corresponding distribution
of gyro-ages calculated using the \citet{kawaler89} age-period-color
relation. The mean age is equal to that derived for the B07 relation,
while the larger $\sigma$ and standard error (3.3\%) reflects primarily
a poorer fit to the M35 I sequence for the late F and early G type
stars, resulting in gyro-ages that are too low for those stars. 

\placefigure{age-distr_kawaler}

%\clearpage
%\clearpage

% -----------------------------------------------------------------------------

\subsection{Improving the I sequence mass-rotation relation using M35}
\label{mass-rotation}

The method of gyro-chronology relies on fitting the I sequence
rotational isochrone, with age as a free parameter, to populations
of cluster stars or to individual field stars in the color-period plane
B07. The functional dependence between stellar color
and rotation period of the isochrone will thus directly affect the
derived gyro-age, and will, if not accurately determined, introduce a
systematic error. It is therefore important to constrain and test
the mass-rotation relation for stars on the I sequence as new data
of sufficiently high quality becomes available.

\begin{figure}[ht!]
\epsscale{1.0}
\plotone{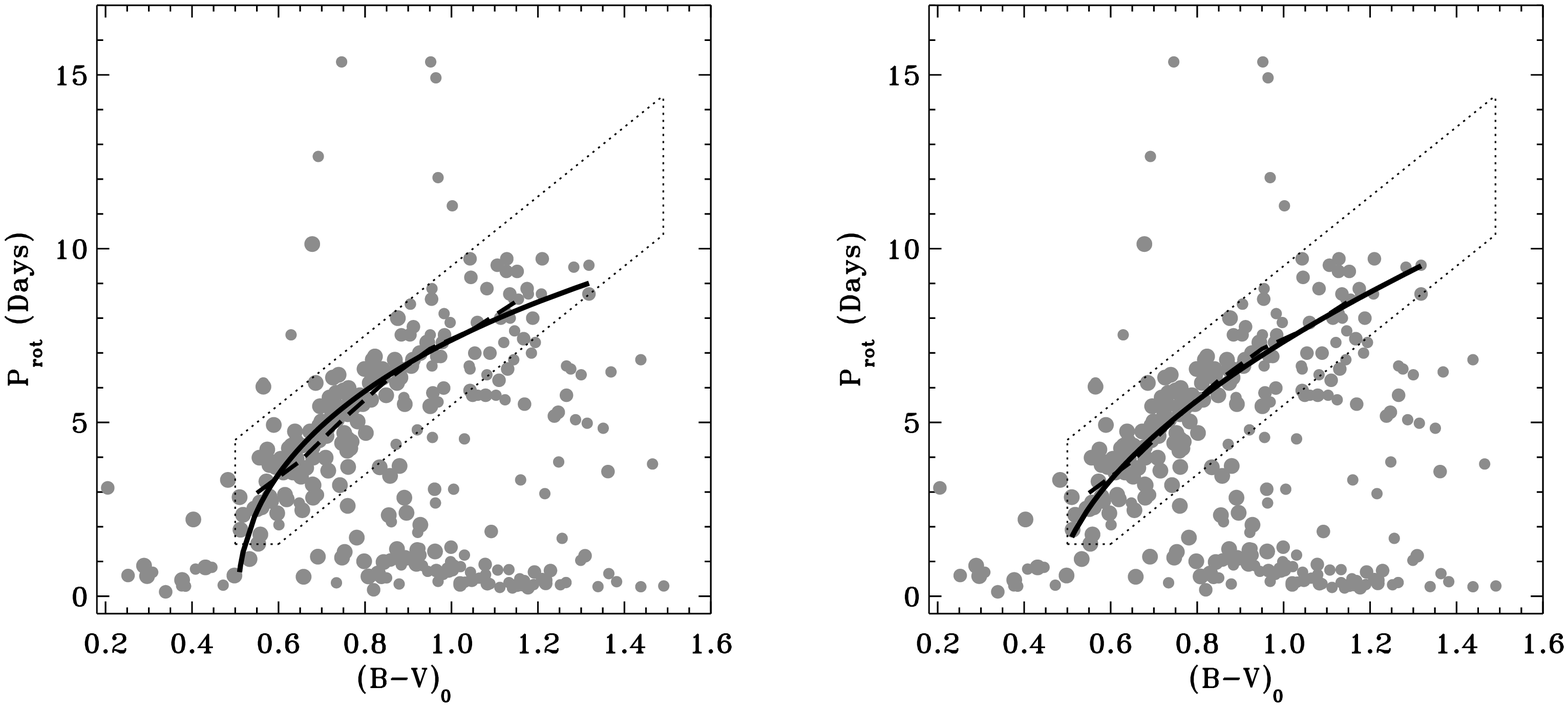}
\caption{The least squares fits (solid curves) of 150\,Myr B03 (left)
and B07 (right) I sequence isochrones (eq. [2] and eq. [3]) to the M35
I sequence. The corresponding new values for the isochrone coefficients
$a$, $b$, $c$, $d$, and $f$ were determined from the fits and are listed
in Table~\ref{coeffs}. The moving average of the rotation periods for the
I sequence stars is also shown as a dashed curve for comparison.
\label{newiso}}
\end{figure}

Our data for M35 are well suited for such a test because
of the rich and well-defined I sequence, the extensive knowledge about
cluster membership, and the independent stellar evolution age for the
cluster. To constrain the color-period relation, we fit equations [2]
and [3] to the M35 I sequence, using the same selection of stars and
the same fitting weights as described in Section~\ref{isochrones}. We
determine all coefficients in equations [2] and [3] for the fixed cluster
age ($t$) of 150\,Myr. The coefficients with $1\sigma$ uncertainties
are listed in Table~\ref{coeffs} and the corresponding rotational
isochrones are shown in Figure~\ref{newiso}. To illustrate how closely
the isochrones trace the selected I sequence stars, Figure~\ref{newiso}
also displays a dashed curve representing the moving average of the rotation
periods along the I sequence.

\placefigure{newiso}

%\clearpage
%\clearpage

We can compare the coefficients derived using the M35 data to those
chosen and/or derived by B03 and B07. Our best fit of the B03 isochrone
confirms the value of 0.5 for the $a$ coefficient chosen (not fitted)
by B03 to best represent the color-period data included in his study.
$a$ is a translational term that determines the color for which the
isochrone gives a period of zero days. For the $b$ coefficient our best
fit give a value of 0.20 compared to the choice of 0.15 by B03. Our
larger value of the $b$ coefficient results in an isochrone with slightly
more curvature.

From our best fit of the B07 rotational isochrone to the M35 I sequence,
we determined a $c$ coefficient of 0.77, equal to the value used by
B07, while our value 0.55 for the $f$ coefficient is smaller than the
value of 0.60 used by B07. In the case of the $c$ and $f$ coefficients,
B07 also determined their values from least squares fitting to the I
sequence stars of several young open clusters. However, for the
translational term $d$, he chose a fixed value of 0.4 to allow for
more blue stars to be fitted. We left the $d$ coefficient as a free
parameter when fitting to the M35 I sequence, and got a value of 0.47.

The new value of 0.47 for $d$ is particularly interesting as it
corresponds to the approximate $B-V$ color for F-type stars at the
transition from a radiative to a convective envelope. This transition
was noted from observations of stellar rotation (known as the break
in the Kraft curve \citep{kraft67}), and is associated with the onset
of effective magnetic wind breaking \citep[e.g.][]{schatzman62}.
The value of 0.47 for the $d$ coefficient therefore suggest that,
for M35, the blue (high-mass) end of the I sequence begins at the
break in the Kraft curve.

%\clearpage
\begin{deluxetable}{cccc}
\tabletypesize{\normalsize}
\tablecaption{New coefficients for the I sequence rotational isochrones
\label{coeffs}}
\tablewidth{0pt}
\tablehead{
\colhead{Isochrone} & \colhead{Coefficient} & \colhead{Value} & \colhead{1$\sigma$ error}\\
%\colhead{} & \colhead{} & \colhead{} & \colhead{}
}
\startdata
B03 & a & 0.507  & 0.005 \\
B03 & b & 0.204  & 0.013 \\
B07 & c & 0.770  & 0.014 \\
B07 & d & 0.472  & 0.027 \\
B07 & f & 0.553  & 0.052 
\enddata
\end{deluxetable}
%\clearpage

% -----------------------------------------------------------------------------

\subsection{Prediction: Tidal Evolution is Responsible for the Unusually
Slow Rotators}					 \label{prediction}

Ten stars fall above the M35 I sequence, and thus rotate unusually
slowly in comparison to other members of M35 with similar masses.
All 10 stars are photometric members of M35 and 2 are also spectroscopic
members. We have no reason to believe that the rotation periods for
these stars are due to aliases in the power spectra, and we note that
a similar pattern is seen in NGC\,3532 with 7 stars located above
the I sequence B03 and in M34 with 6 stars above the I sequence
\citep{mms08b}.

What causes the rotational evolution of these stars to deviate
significantly from that of most similar-color stars in M35? We propose
here that tidal interactions with a close stellar companion has
acted to partially or fully synchronize the stellar spins of these
stars to the orbital motions, and that such tidal synchronization
is responsible for their slower-than-expected rotation. We thus
predict that these 10 stars are the primary stars in binaries with
periods of $\sim$10-15 days.

This proposition finds support from the star of M35 located in the
color-period diagram at $(B-V)_0 = 0.68$ and $P_{rot} = 10.13$ days.
This star is the primary in a circular binary with an orbital period
of 10.33 days. The rotation of this star has been synchronized to the
orbital motion of the companion \citep{mms06}, forcing it to rotate
more slowly than stars of similar mass.
In addition to the spectroscopic binary, 3 of the remaining
9 slow rotators are photometric binaries. Spectroscopic observations
has begun of those stars and of the remaining 5 stars as of fall
2007 to determine their status as binary or single stars, and in
the case of binarity, the degree of tidal evolution.

% -----------------------------------------------------------------------------

\subsection{Stellar angular momentum evolution near the ZAMS}
\label{hypothesis}

The trend in Figure~\ref{fractions} of an increasing fraction of C
sequence (and gap) stars for younger cluster populations leads 
naturally to the suggestion that most, if not all, late-type stars
pass through a phase of rapid rotation (the C sequence) at the ZAMS.
We note that even should this be the case, an {\it observed} C sequence
fraction of 1 is not expected for even the youngest coeval populations,
as stars of different masses will reach the ZAMS, and thus hypothetically
the C sequence, at different times. For example, late-F stars will be
the first to arrive at/on the ZAMS and C sequence, and leave it before
the arrival of G and K type stars.

The color-period diagrams for the youngest stellar populations
presented in Figure~\ref{fractions} (see also Figure 1 in B03)
show that most stars lay either on the C sequence or in the gap.
B03 finds only 25\% of the stars at 30 Myr to be on the I sequence.
In fact the I sequence is not clearly identifiable at this age,
and the stars identified as being on the I sequence are early-type
rapid rotators near the intersection of the two sequences.

By 30 Myr very few of the cluster members have rotation periods longer
than 5 days. This is in marked contrast to fractions of $\sim$60\% and
$\sim$40\%, respectively, for such slow rotators in the PMS populations
of the Orion Nebula cluster (ONC) and NGC\,2264 \citep[see][]{hem+07}.
The difference in the numbers of slowly rotating stars pre- and post-ZAMS,
suggests that most, if not all, of the stars rotating slowly at $\sim$1-3\,Myr,
spin up as they evolve onto the ZAMS. Such spin-up may have been observed.
Comparison of the rotation period distributions for stars in the $\sim$1\,Myr
ONC and the $\sim$2-3\,Myr NGC\,2264 \citep[][and references therein]{hem+07}
shows a spin-up with time by a factor $\sim$2, presumably due to conservation
of angular momentum as the stars contract on the PMS. On the other hand,
the distribution of a smaller sample of rotation periods in the $\sim$2-3\,Myr
IC\,348 \citep{nhr+06} does not show similar evidence for spin-up when
compared to the ONC.

From the point of view of modeling stellar angular momentum evolution,
we emphasize the narrowness of the C sequence, with all rotation periods
between 0.5 days and 1.5 days. We suggest that the broad distribution of
rotation period among solar-like stars in the PMS populations must collapse
into a narrow C sequence of similar rotation periods independent of mass.
Indeed, we suggest that in the two 30\,Myr clusters of B03 (Figure 1),
the gap stars with $(B-V)_{0} \ga 0.9$ may in fact be evolving {\it toward}
the C sequence, and point out that in the 50\,Myr clusters in B03, mostly C
sequence stars are observed redward of $(B-V)_{0} \simeq 0.9$.

% *****************************************************************************

\section{SUMMARY AND CONCLUSIONS} \label{conclusions}

We present the results of an extensive time-series photometric survey
over $\sim$5 months of late-type members in the 150\,Myr open cluster
M35 (NGC\,2168). We have obtained photometric light curves for 14022
stars with $12 \la V \la 19.5$ over a $40\arcmin \times 40\arcmin$ field
centered on M35. We have determined the rotation periods for 441 stars.
Cluster membership and binarity for stars with rotation periods are
determined from the results of a decade long spectroscopic survey in M35.
Of the 441 rotators 310 stars are radial-velocity and/or photometric
members of M35. 

With an age slightly older than the Pleiades but with a much larger
population of late-type stars, M35 is particularly interesting for
studying stellar rotational evolution during this active phase of
angular momentum evolution between the ZAMS and the age of the Hyades.
The rotation periods of the 310 late-type members span over two
orders of magnitude from 0.1 day ($\ga50\%$ of their breakup velocities),
up to $\sim$15 days. A drop-off in the period distribution is found
at $\sim$10 days, well below the upper limit of our period search.
The $\sim$10-day cutoff may represent a physical upper limit on the
rotation-period distribution at 150 Myr. However, it is also possible
that detecting more slowly rotating stars in M35 will require higher
photometric precision or higher resolution spectroscopic observations.

We find in the phased light curves for almost all stars with measured
rotation periods that the long-baseline ($\sim$5 months), low-frequency
(1/night) photometric measurements match the short-baseline
(16 nights), high-frequency ($\sim$ 1/hours) measurements in both
phase, shape, and amplitude. Further tests on a subset of stars
show that the same rotation periods are derived from the short-
and long-term data to within 1\%. This stability in the modulation
of the stellar brightness suggest a similar stability in the
configuration, size, and number of starspots.

In the color-period plane, the 310 M35 rotators reveal striking
dependencies between surface rotation period and stellar color
(mass). More than 75\% of the stars lay along two distinct sequences
in the color-period diagram, apparently representing two different
states in their rotational evolution. Similar sequences
were identified by \citet{barnes03a} for stars
in other clusters. Comparison between M35 and these clusters of the
locations of the sequences in color-period diagram, as well as the
relative numbers of stars on each, support for the idea (proposed by
\citet{barnes03a}) that stars evolve from one sequence (C) to the
other sequence (I) at a rate that is inversely proportional to the
stellar mass.

We determine from the M35 color-period diagram that the characteristic
exponential timescale for rotational evolution off the C sequence and
onto the I sequence is $\sim$60 Myr and $\sim$140 Myr for G and K dwarfs,
respectively. These timescales may offer valuable constraints on the
rates of internal and external angular momentum transport and on the
evolution rates of stellar dynamos in late-type stars of different masses.

From the emerging trend (supported by M35) of an increasing relative
fraction of rapidly rotating C sequence stars with decreasing population
age, we propose the hypothesis that most, if not all, late-type stars
pass through a phase of rapid rotation (C sequence) on the ZAMS.
By conjecture, there may not be a need for a direct connection between
slowly rotating stars observed in the early PMS and slowly rotating
stars at $\sim$100\,Myr post the ZAMS. Such a connection has often
been assumed and set as a constraint on models of stellar angular
momentum evolution, motivating the introduction of mechanisms to
prevent slowly rotating PMS stars from spinning up as they evolve
onto the main-sequence.

By comparison with measured rotation periods in the Hyades, we
put to the test the empirical Skumanich $\sqrt{t}$ time-dependence
on the stellar rotation period for G dwarfs. By reducing the Hyades
rotation periods by a factor $\sqrt{Age_{Hyades}/Age_{M35}}$ we find
that the $\sqrt{t}$ law accounts very well for the rotational evolution
of G dwarfs between M35 and the Hyades, whereas among the K dwarfs
the $\sqrt{t}$ time-dependence predicts a spin-down rate that is
faster than observed between M35 and the Hyades.

We find that the heuristic rotational isochrones proposed by
\citet{barnes03a} and \citet{barnes07} match the location of M35
I and C sequences using the independently determined stellar evolution
isochrone age for M35. A non-linear least-squares fit of the rotational
isochrones to the M35 I sequence sets the cluster's gyro-age to 134\,Myr
with a formal $1\sigma$ uncertainty of 3 Myr. We use the age-period-color
relations by \citet{barnes03a}, \citet{barnes07}, and \citet{kawaler89},
to calculate the distributions of gyro-ages for the M35 I sequence
stars. The mean gyro-ages have standard errors of order 3\% and agree
well with the $\sim$150\,Myr age derived for the cluster using the
isochrone method.
These results suggest that a well-populated color-period diagram,
cleaned for non-members, in combination with rotational isochrones,
can provide a precise age estimate that is consistent with the age derived
from isochrone fitting in the CMD. We also use the M35 I sequence to improve
the coefficients for the color-dependence of the rotational isochrones.

Finally, to explain the ten M35 stars rotating with rates that are
unusually slow compared to similar stars in the cluster, we propose
that tidal synchronization in binary stars with orbital periods of
order 10-15 days is responsible. Two of the 10 stars have already been
found to be primary stars in tidally evolved spectroscopic binaries,
while 3 other stars are photometric binaries. Accordingly, we predict
that the remaining stars are also primary stars in spectroscopic binaries
with orbital periods of $\sim$10-15 days.

% *****************************************************************************

\acknowledgments

We wish to thank the University of Wisconsin - Madison Astronomy
Department and NOAO for the time granted on the WIYN 0.9m and
3.5m telescopes. We express our appreciation to the site managers
and support staff at both telescopes for their exceptional and
friendly support. We are thankful to all observers in the WIYN
0.9m consortium who provided us with high-quality data through
the queue-scheduled observing program. This work has been supported
by NSF grant AST-0406615 to the UW-Madison, a Ph.D fellowship from
the Danish Research Academy to S.M., partial support to S.M. from
the Kepler mission via NASA Cooperative Agreement NCC2-1390, and
by the Cottrell Scholarship from the Research Corporation to K.G.S.

\clearpage

% *****************************************************************************
% 	A	P	P	E	N	D	I	X


\appendix

\section{PHASED LIGHT CURVES}

This appendix presents the light curves for the stars in the field of M35
for which we measured rotation periods. In the printed journal,
Figure~\ref{f15} below show examples of our light curves and light
curve plots. Phased light curves for all 441 stars can be found in the
electronic edition of the Journal. The light curves have
been divided into 3 groups according to the amplitude of the photometric
variation. For each group the light curves are sorted by the rotation period
and are presented with the same $\delta V$ range on the ordinate. The group
of stars with the largest photometric variability are shown first.

For each star we plot the data from the high-frequency survey (December 2002)
as black symbols and data from the low-frequency survey (October 2002 through
March 2003) as grey symbols. A running ID number corresponding to the ID
number in Table 1 Appendix B is given in the upper left hand corner in each
plot. The period to which the data are phased (the rotation period listed in
Table 1 as $P_{rot}$) is given in the upper right corner. The 2-5 letter code
in the lower right corner informs about the stars membership status. The
codes have the following meaning: Photometric Member (PM; described in
Section~\ref{phot}), Photometric Non-Member (PNM), Photometric and
Spectroscopic Member (PSM), Photometric and Proper-motion Member (PPM),
and Photometric Member but Spectroscopic Non-Member (PMSNM). For each star
a horizontal grey line in each plot mark $\delta V = 0.0$ and a vertical
grey line marks a phase of 1.0.

\notetoeditor{The first panel (f15a.eps) of Figure 14 could be shrunk
slightly and rotated 90 degrees to fit below the text above}

\placefigure{f15}

\clearpage
\pagestyle{empty}
\begin{figure}
\epsscale{.92}
\plotone{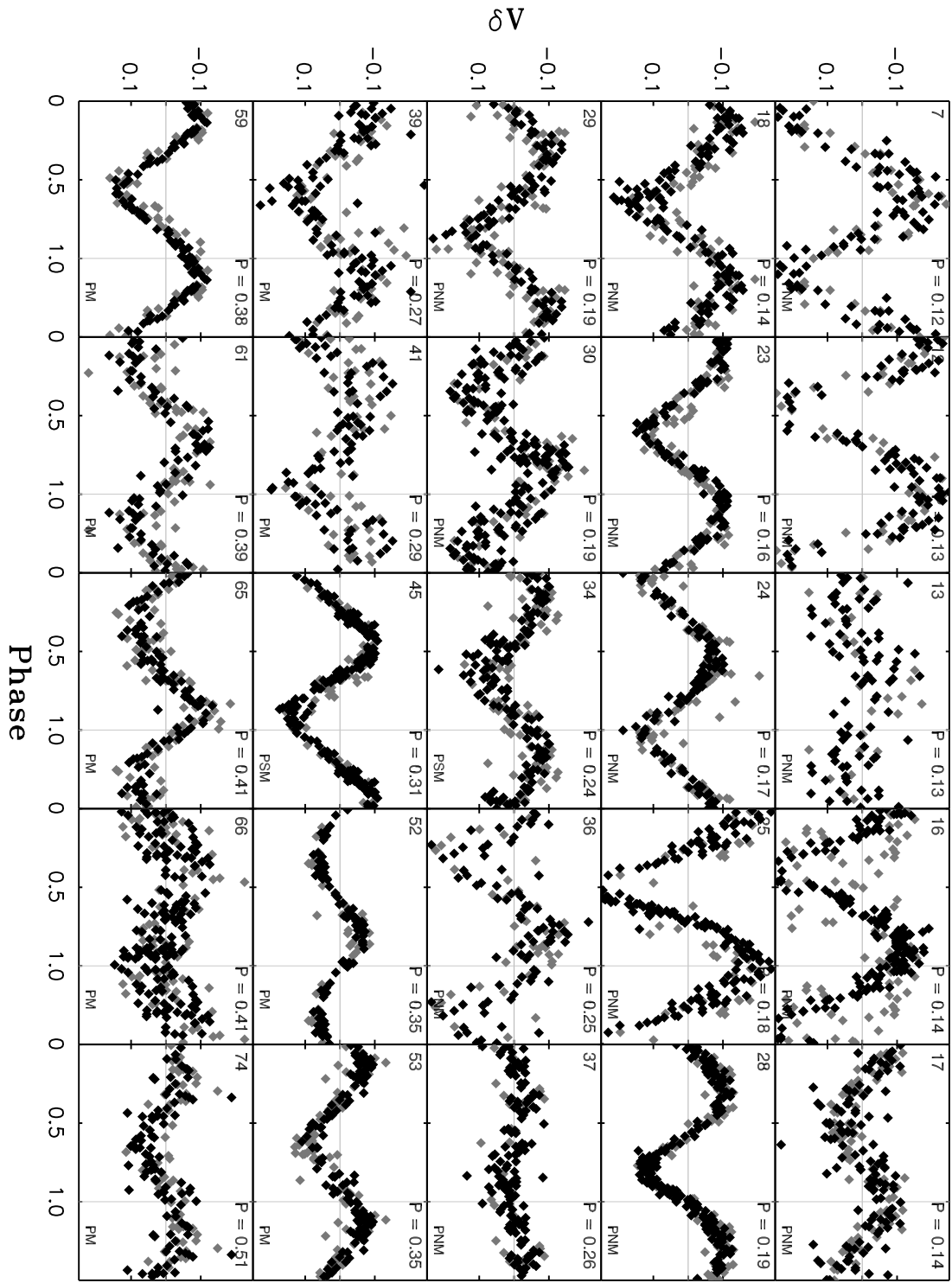}
\caption{Phased light curves for stars with measured rotation periods.
%This figure is intended to show examples of our light curves and light
%curve plots. Phased light curves for all 441 stars can be found in the
%electronic edition of the Journal.
\label{f15}}
\end{figure}
\clearpage
{\plotone{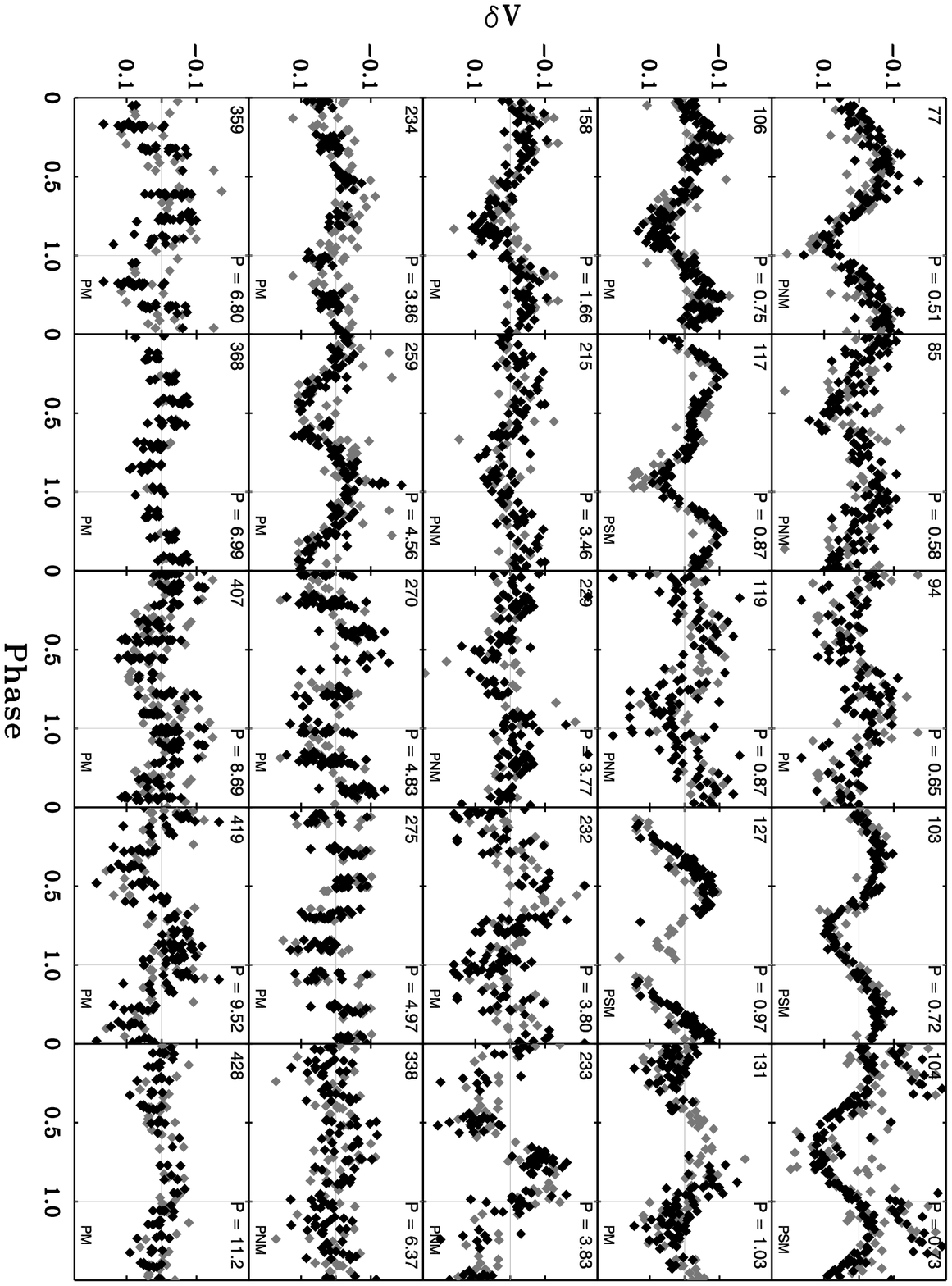}}\\
\centerline{Fig. 14. --- Continued.}
\clearpage
{\plotone{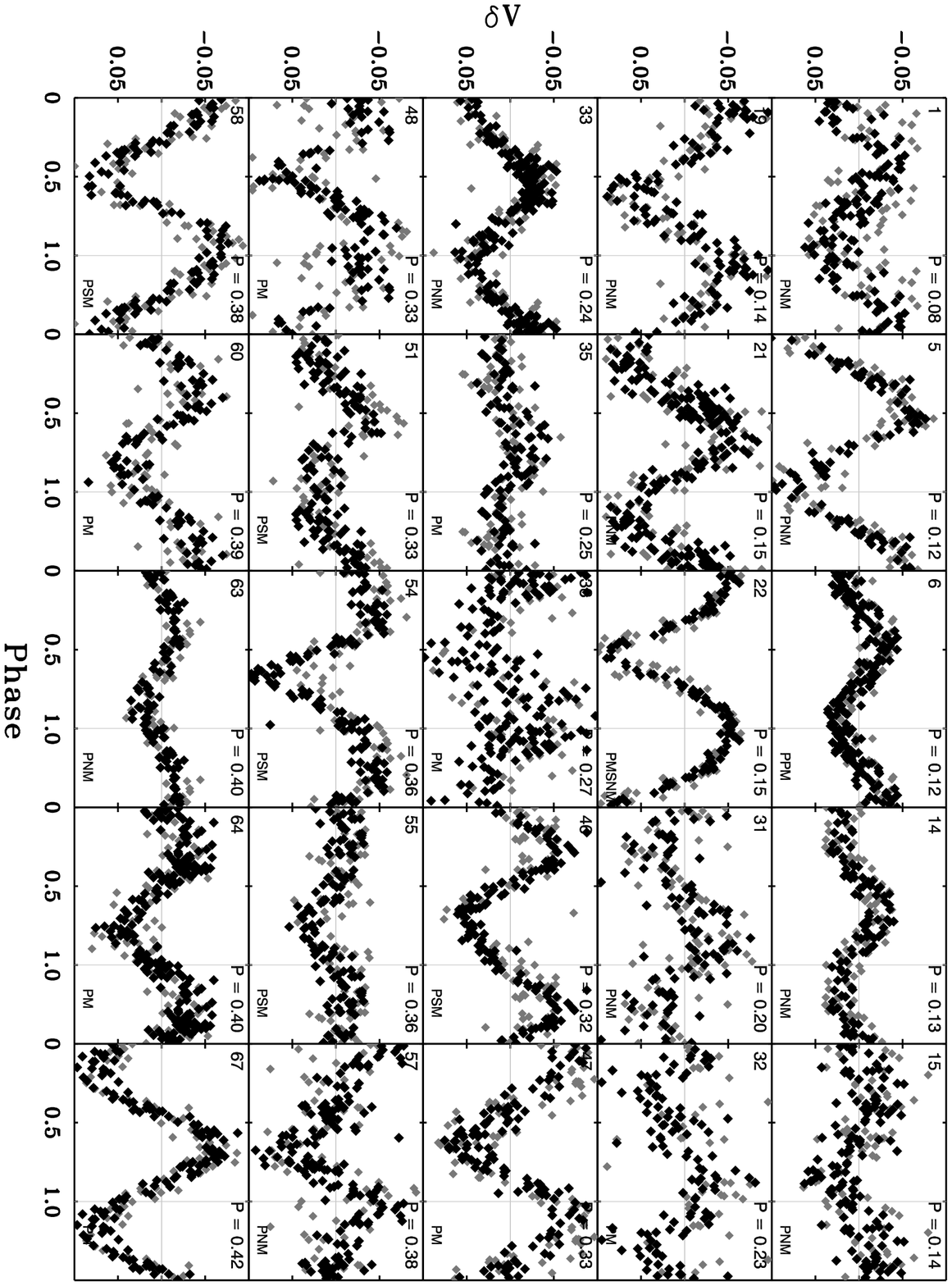}}\\
\centerline{Fig. 14. --- Continued.}
\clearpage
{\plotone{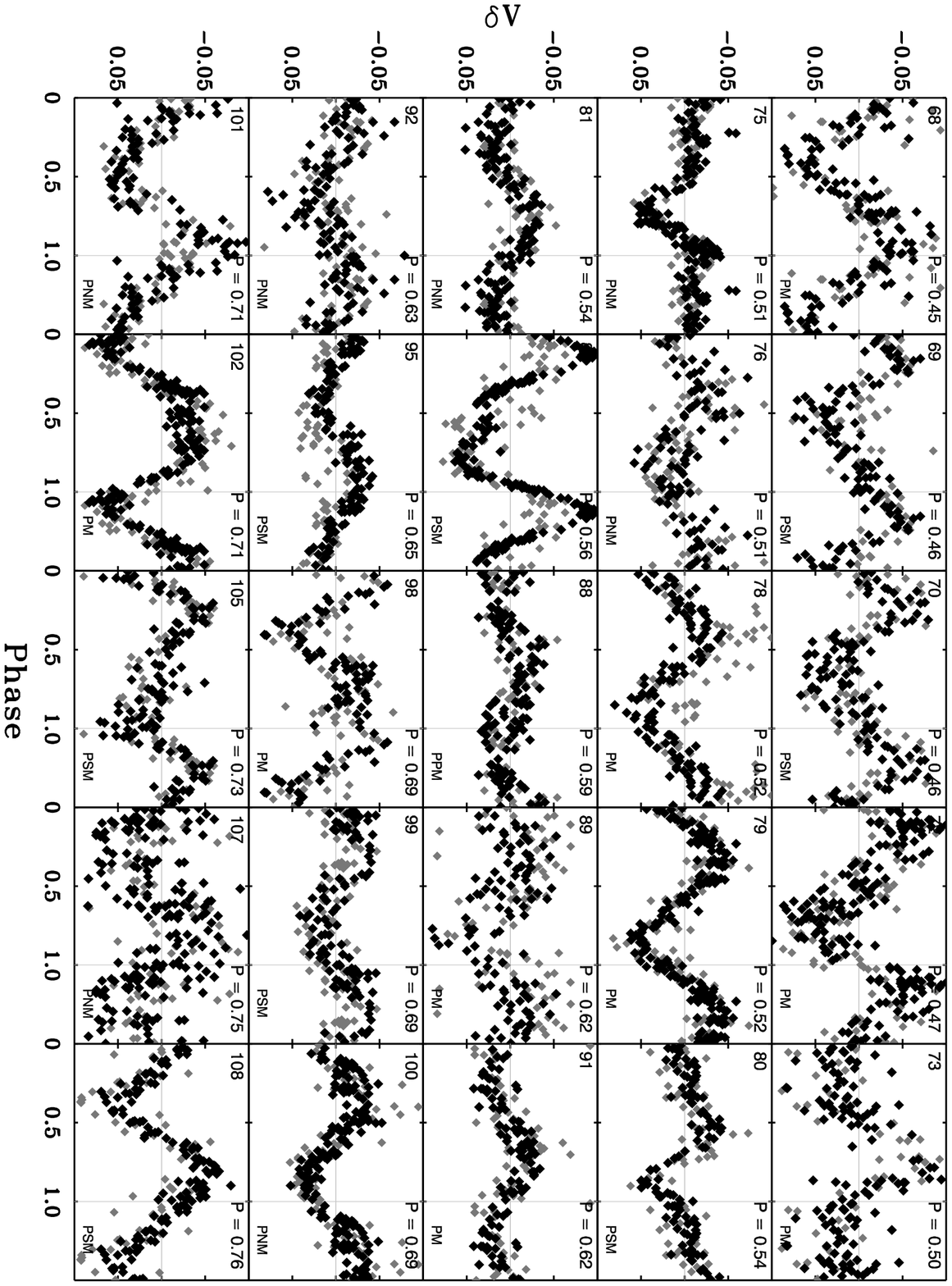}}\\
\centerline{Fig. 14. --- Continued.}
\clearpage
{\plotone{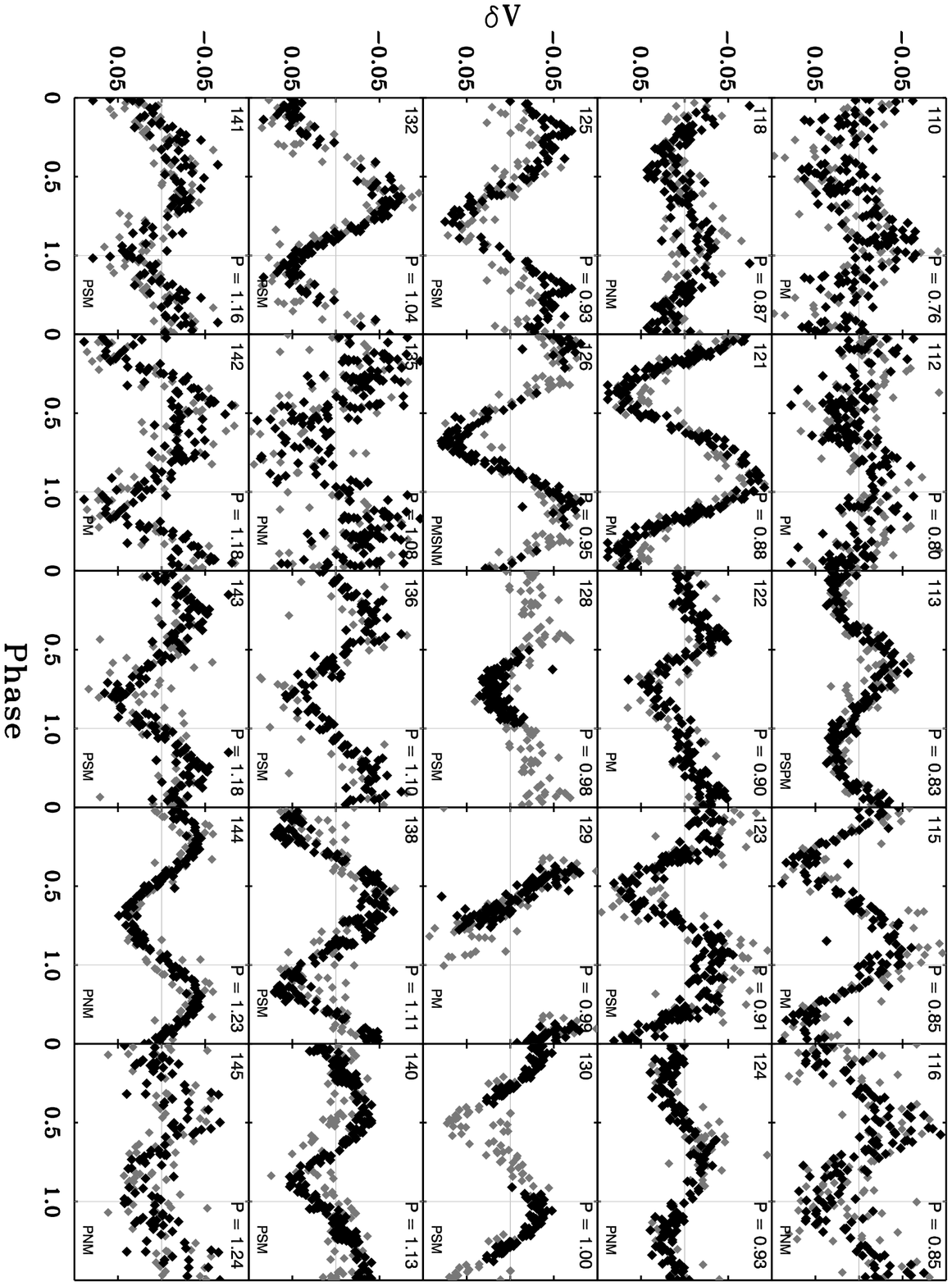}}\\
\centerline{Fig. 14. --- Continued.}
\clearpage
{\plotone{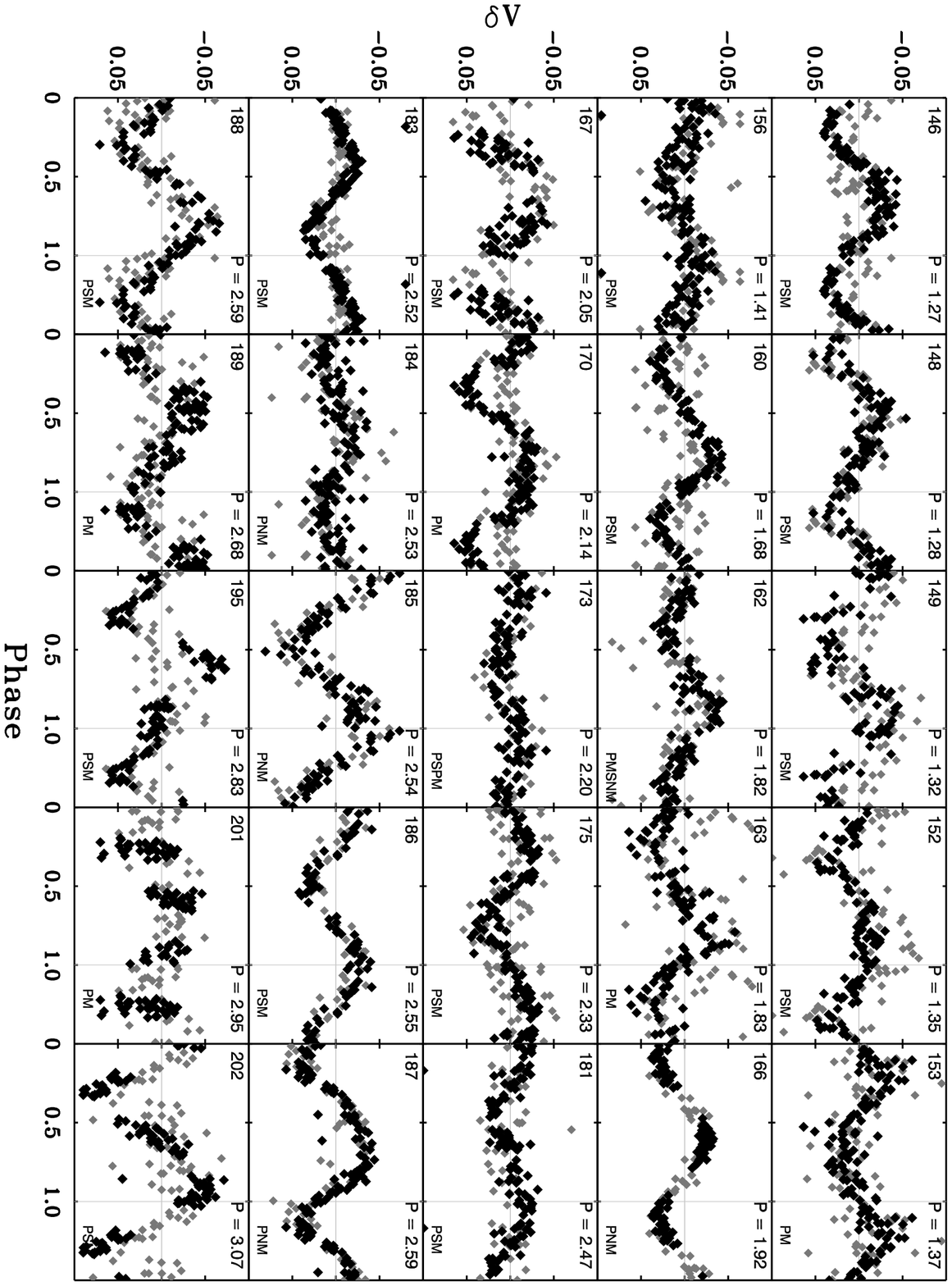}}\\
\centerline{Fig. 14. --- Continued.}
\clearpage
{\plotone{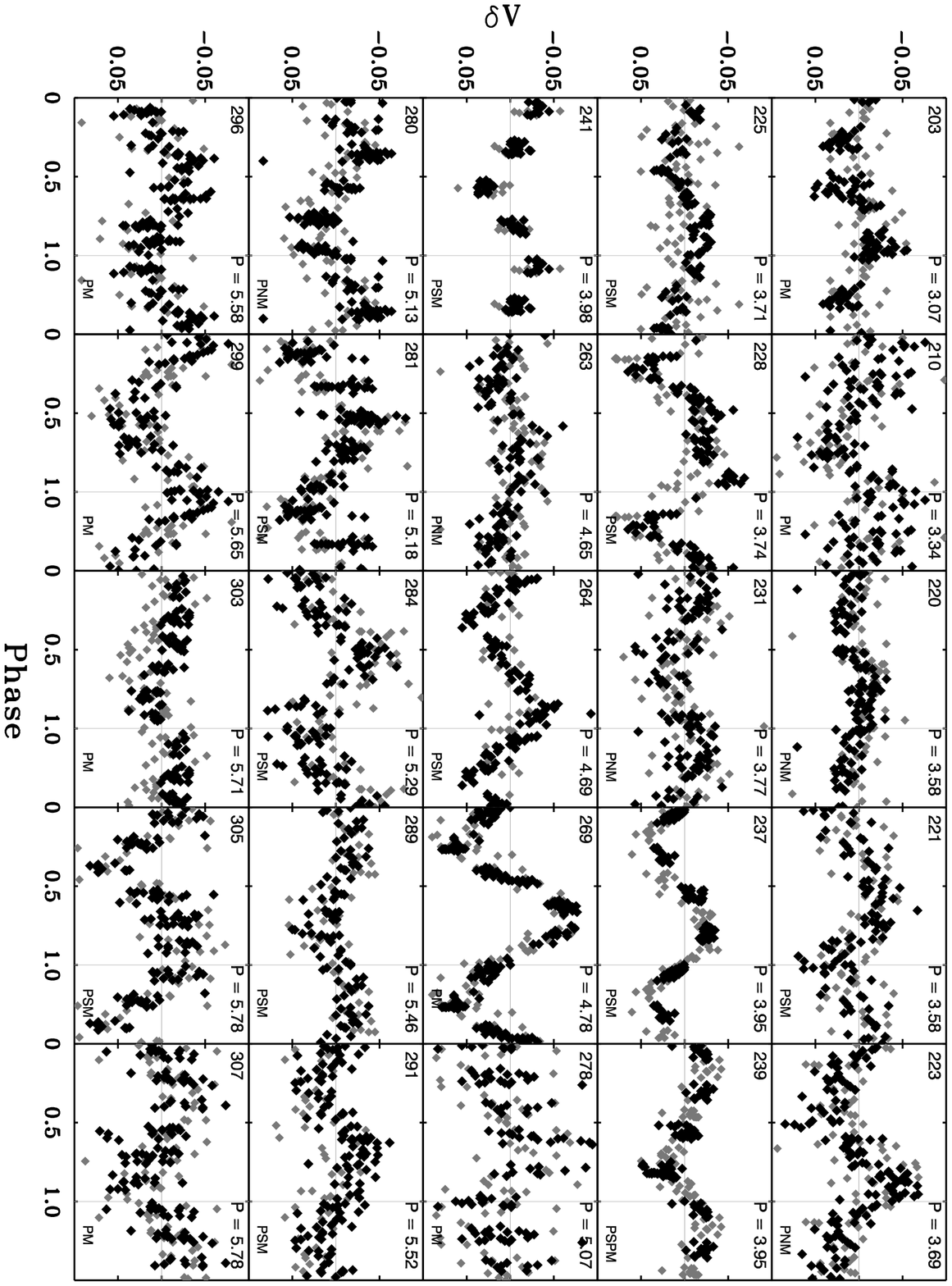}}\\
\centerline{Fig. 14. --- Continued.}
\clearpage
{\plotone{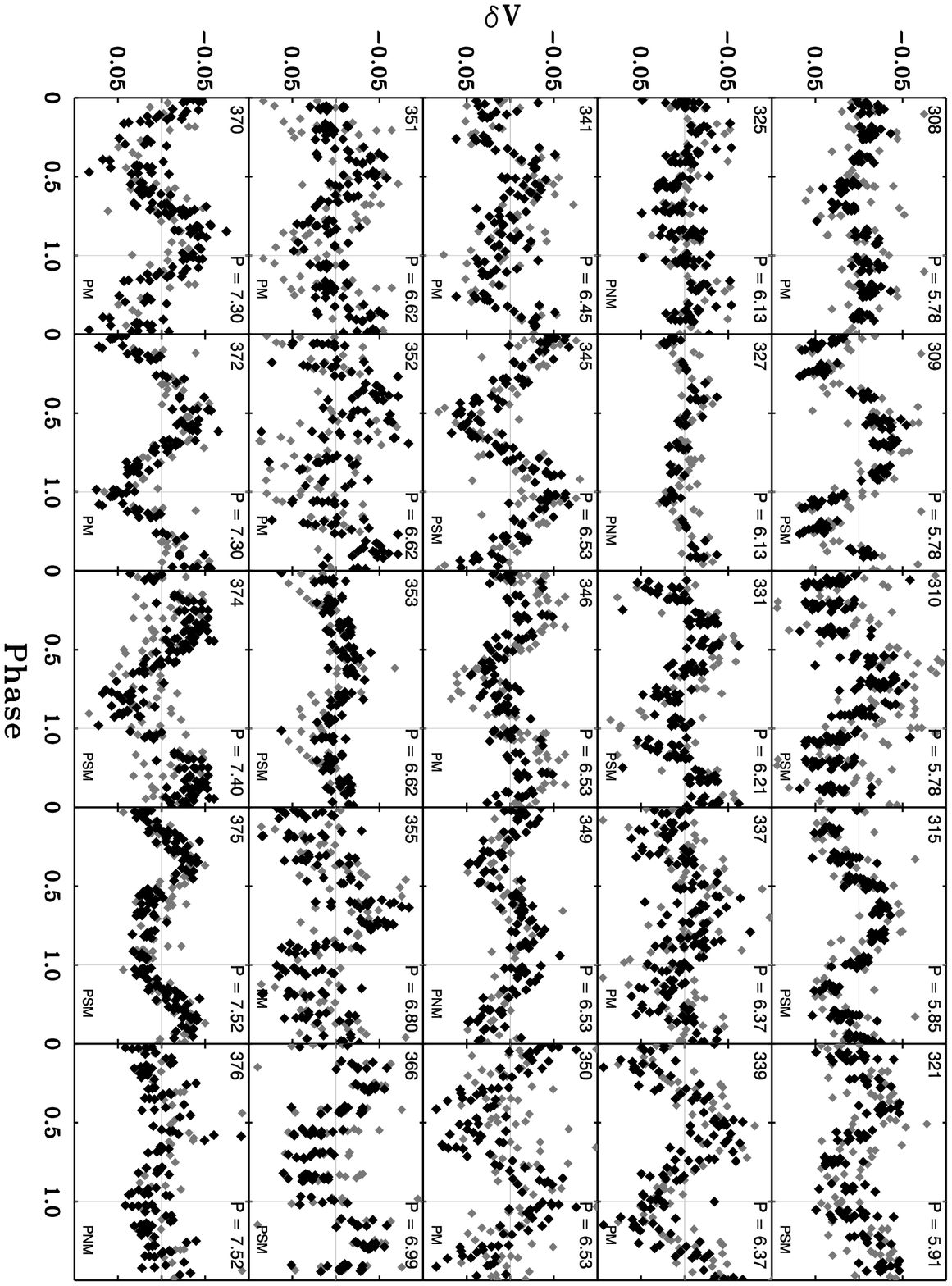}}\\
\centerline{Fig. 14. --- Continued.}
\clearpage
{\plotone{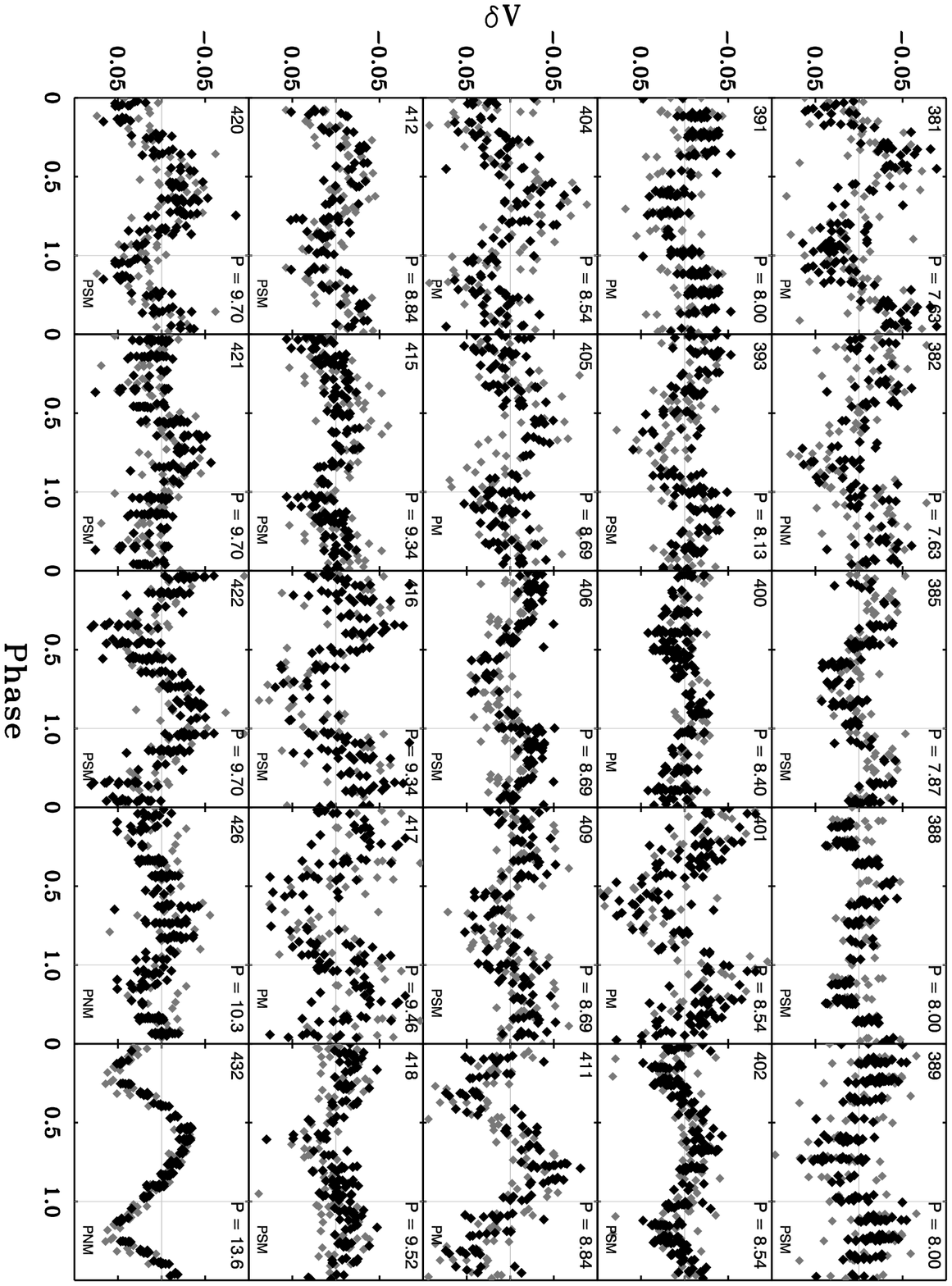}}\\
\centerline{Fig. 14. --- Continued.}
\clearpage
{\plotone{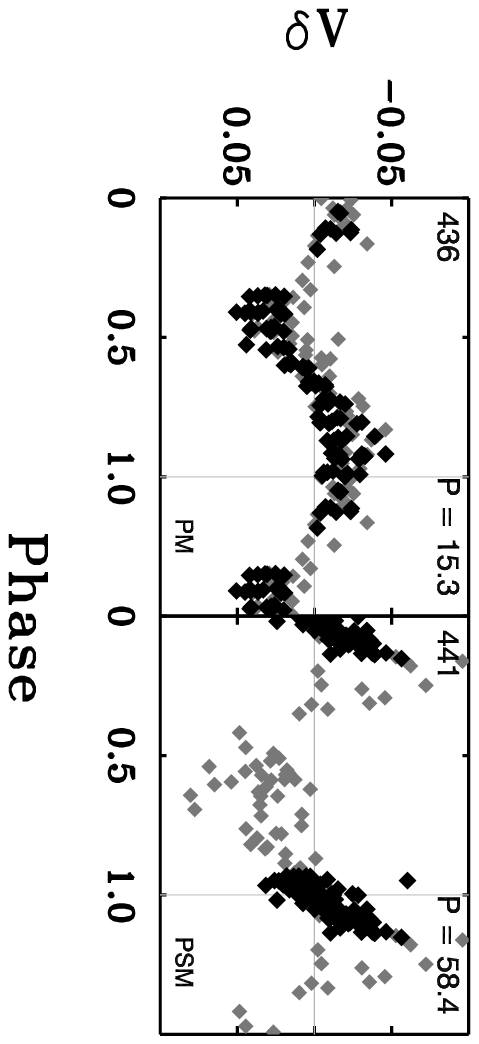}}\\
\centerline{Fig. 14. --- Continued.}
\clearpage
{\plotone{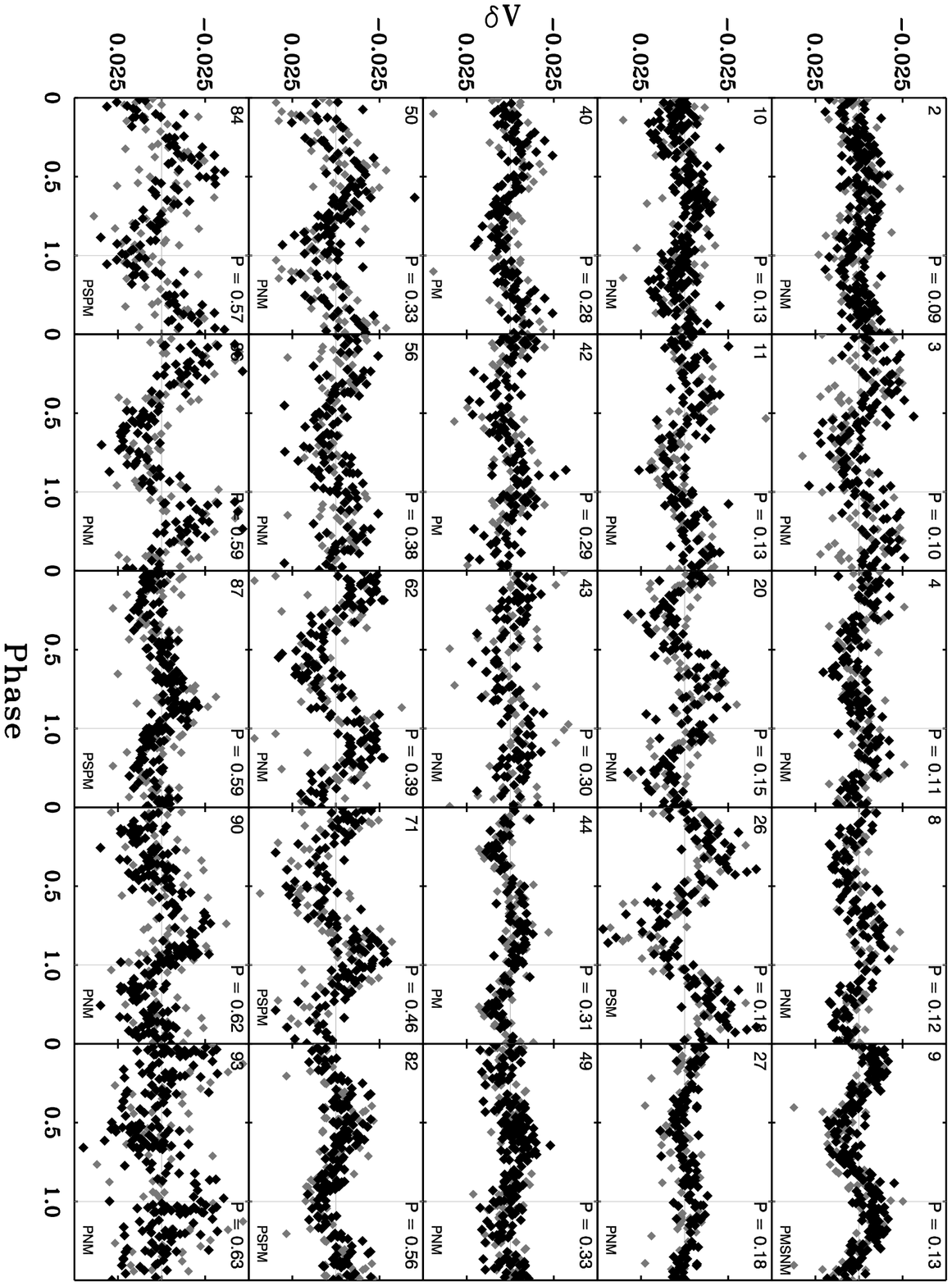}}\\
\centerline{Fig. 14. --- Continued.}
\clearpage
{\plotone{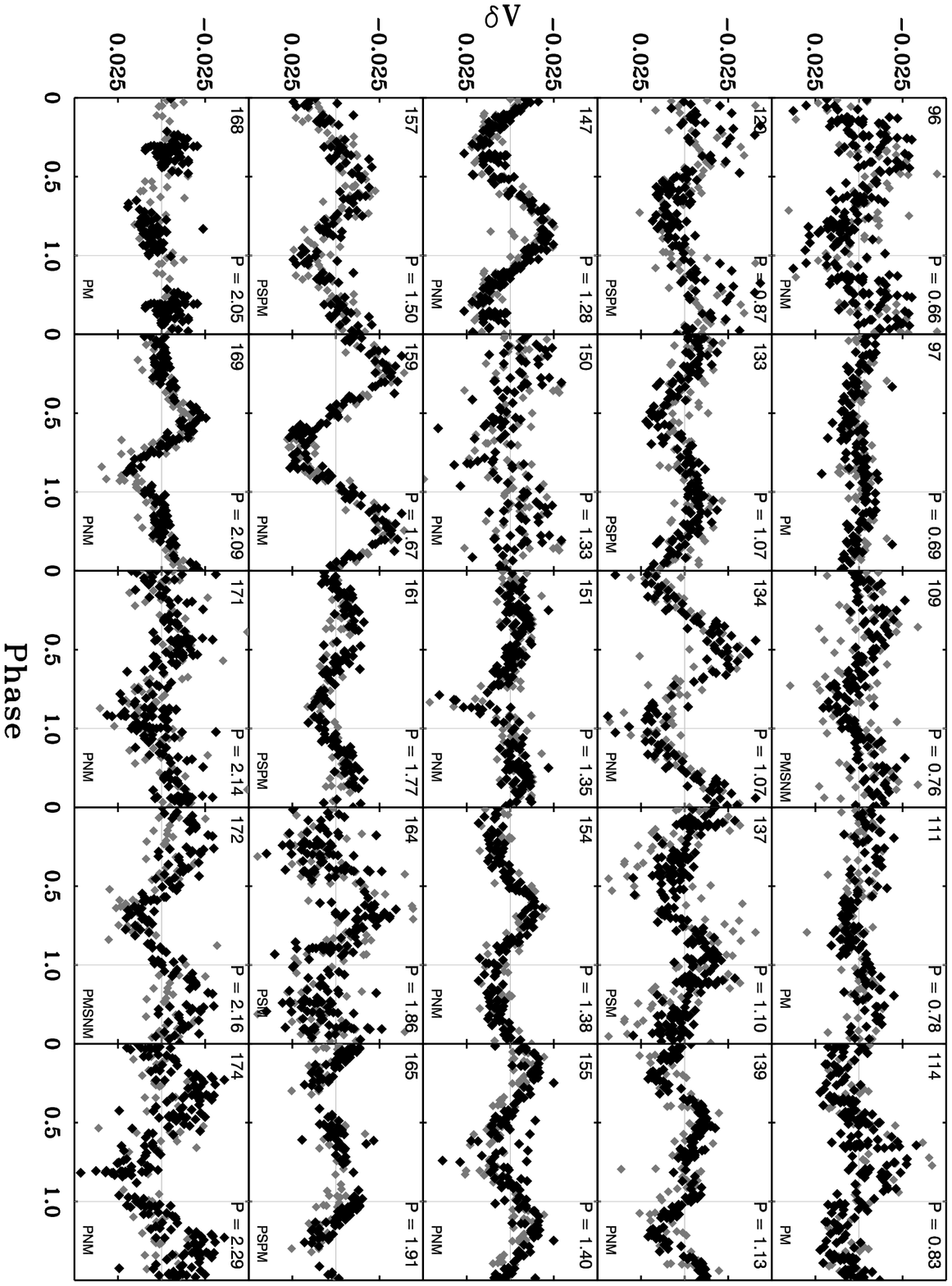}}\\
\centerline{Fig. 14. --- Continued.}
\clearpage
{\plotone{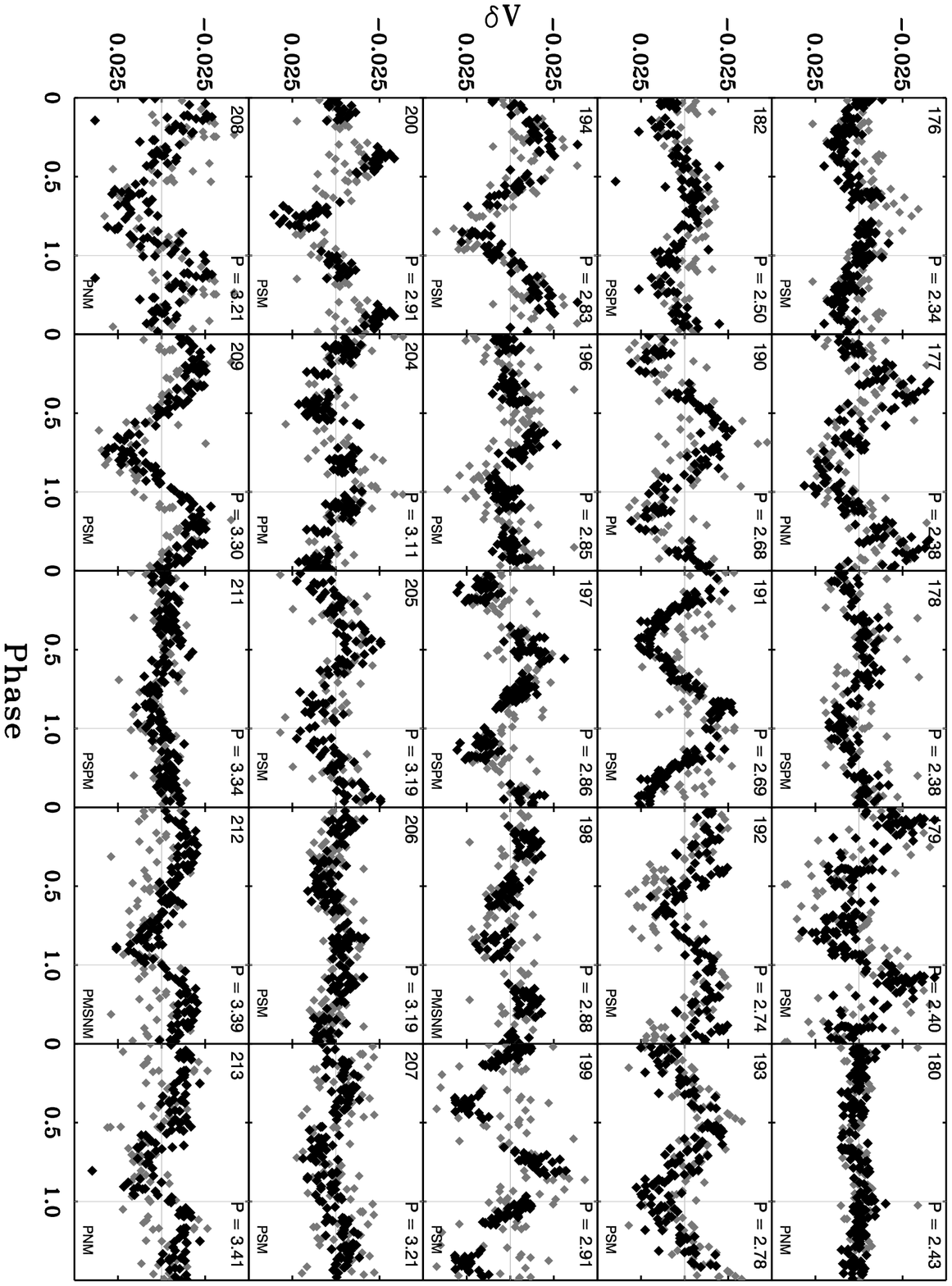}}\\
\centerline{Fig. 14. --- Continued.}
\clearpage
{\plotone{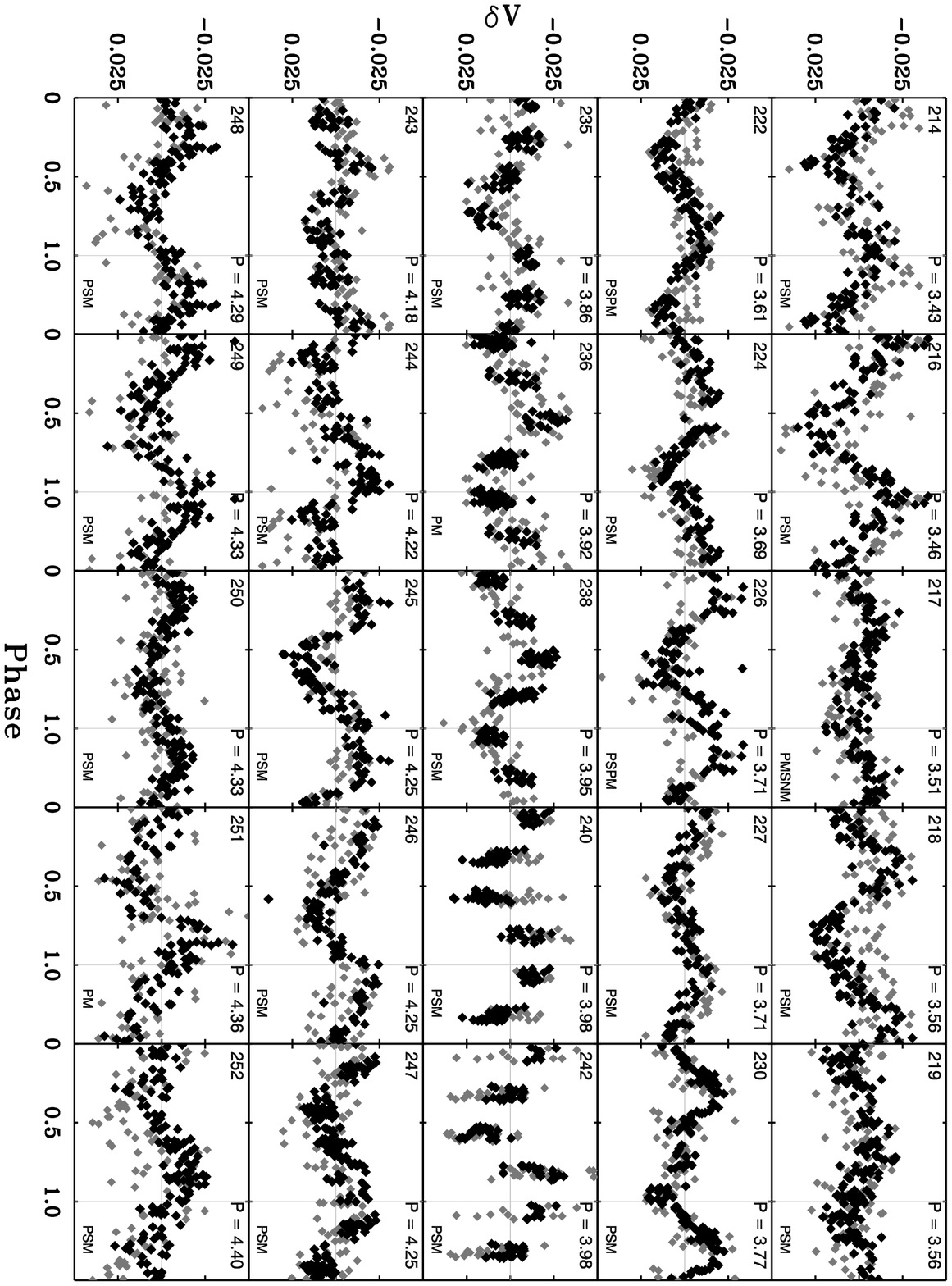}}\\
\centerline{Fig. 14. --- Continued.}
\clearpage
{\plotone{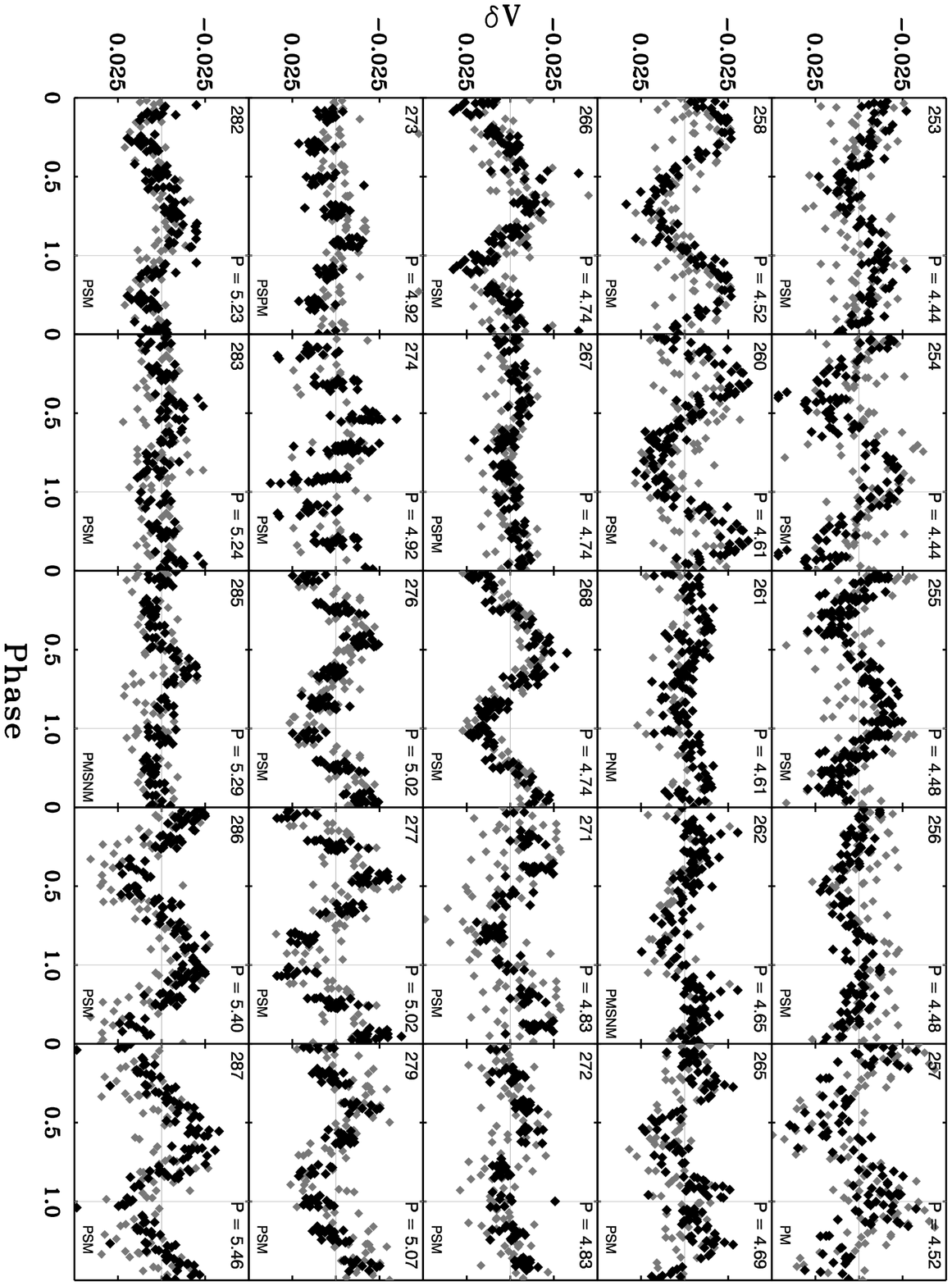}}\\
\centerline{Fig. 14. --- Continued.}
\clearpage
{\plotone{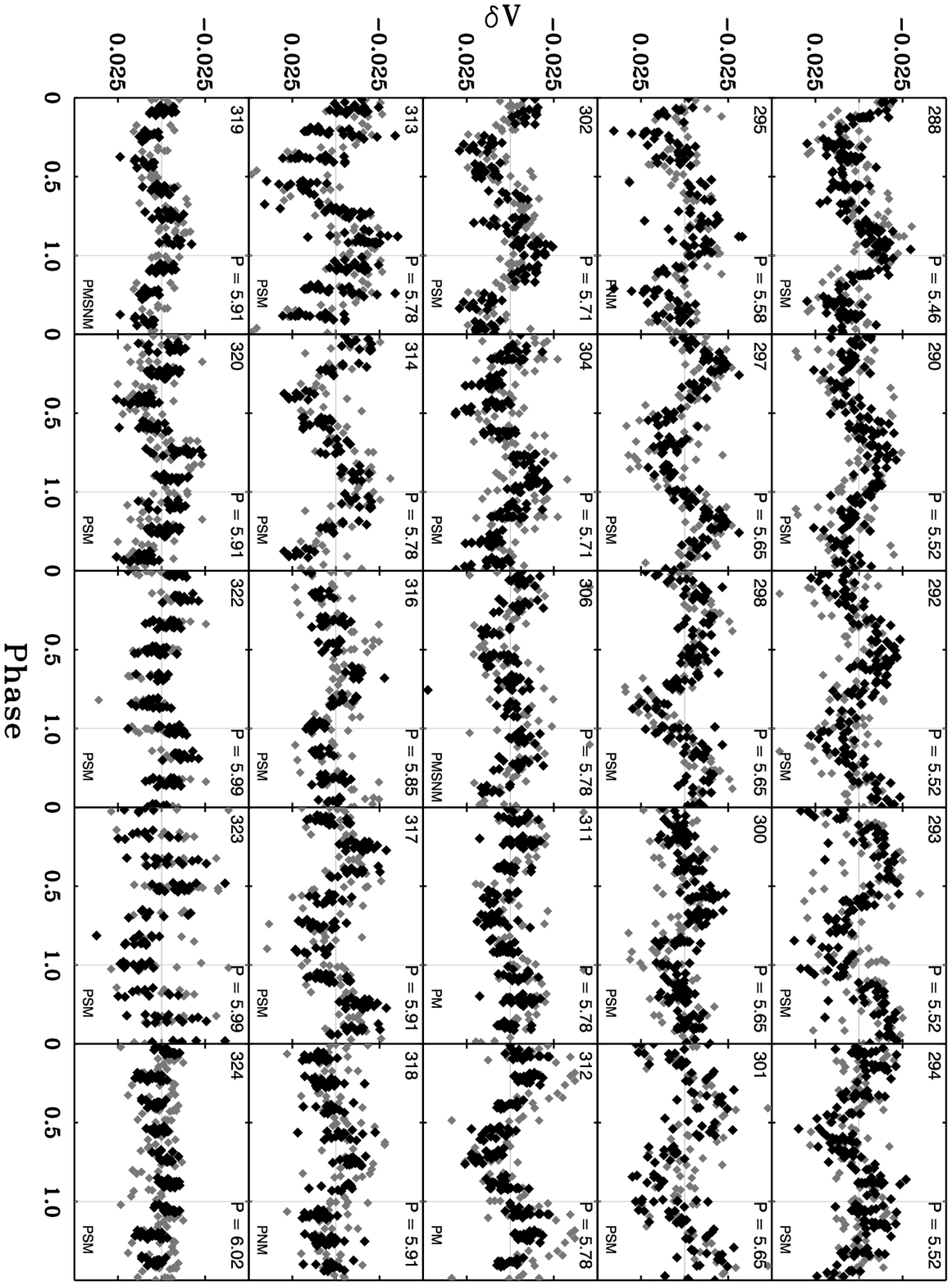}}\\
\centerline{Fig. 14. --- Continued.}
\clearpage
{\plotone{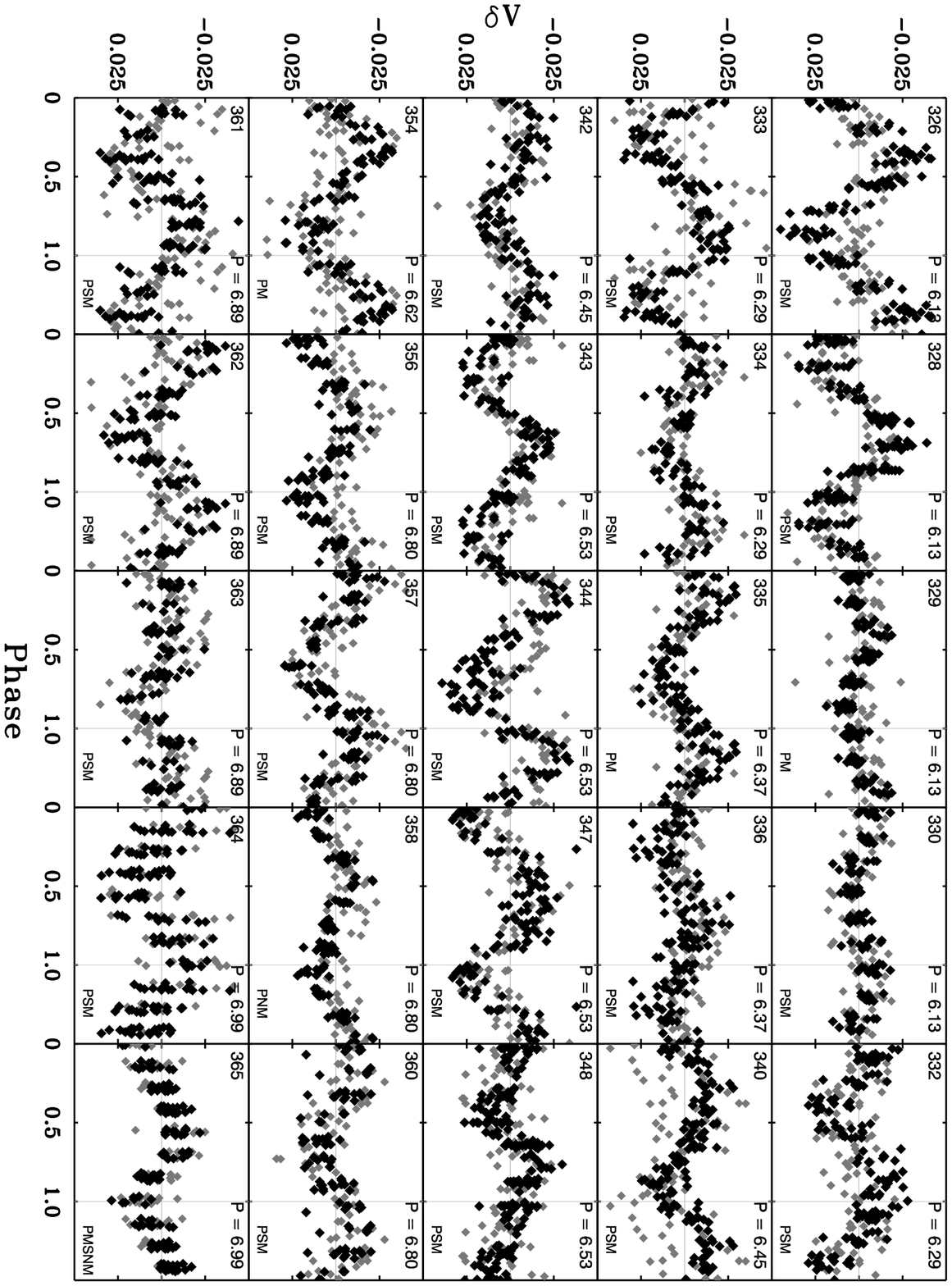}}\\
\centerline{Fig. 14. --- Continued.}
\clearpage
{\plotone{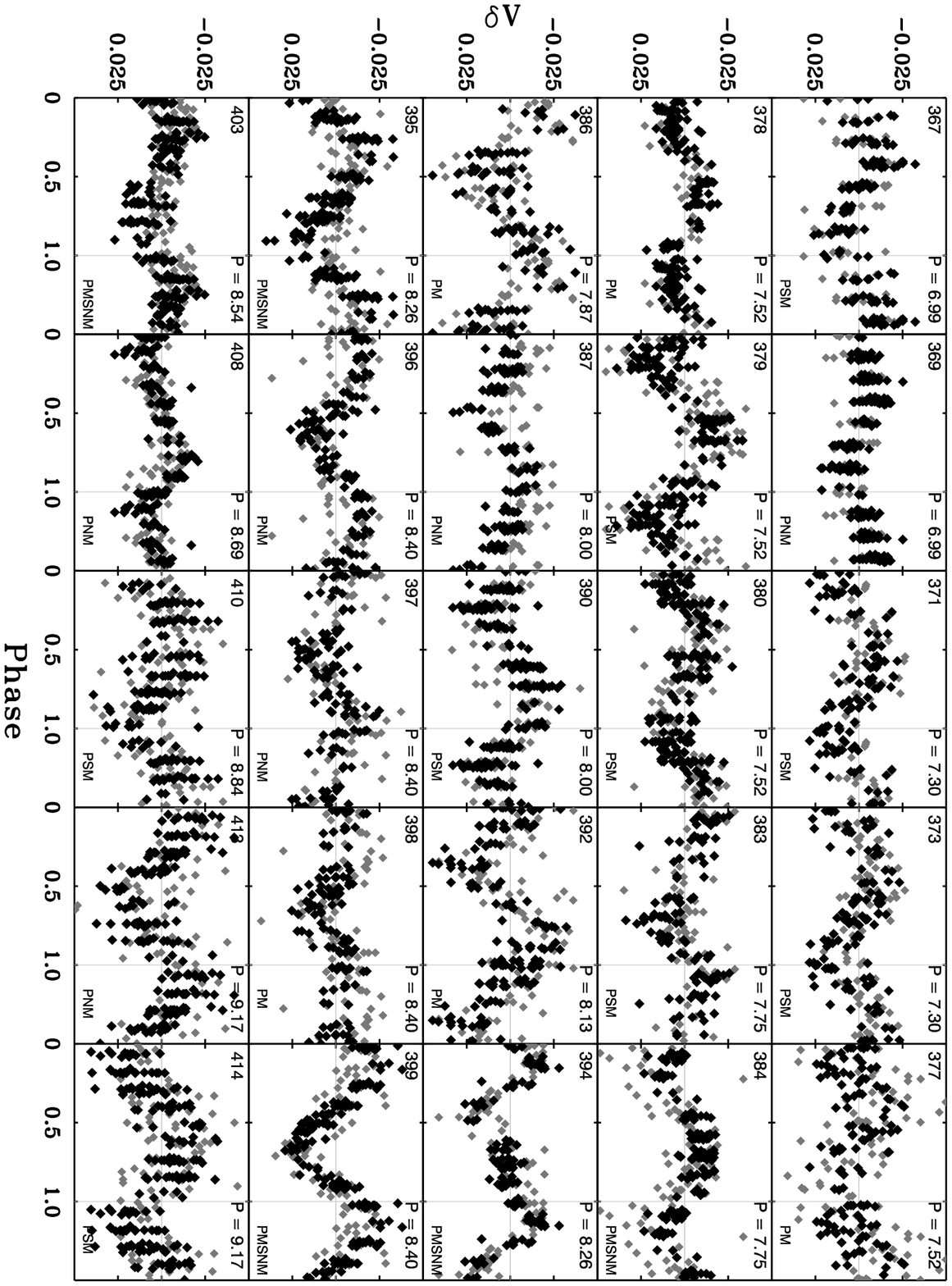}}\\
\centerline{Fig. 14. --- Continued.}
\clearpage
{\plotone{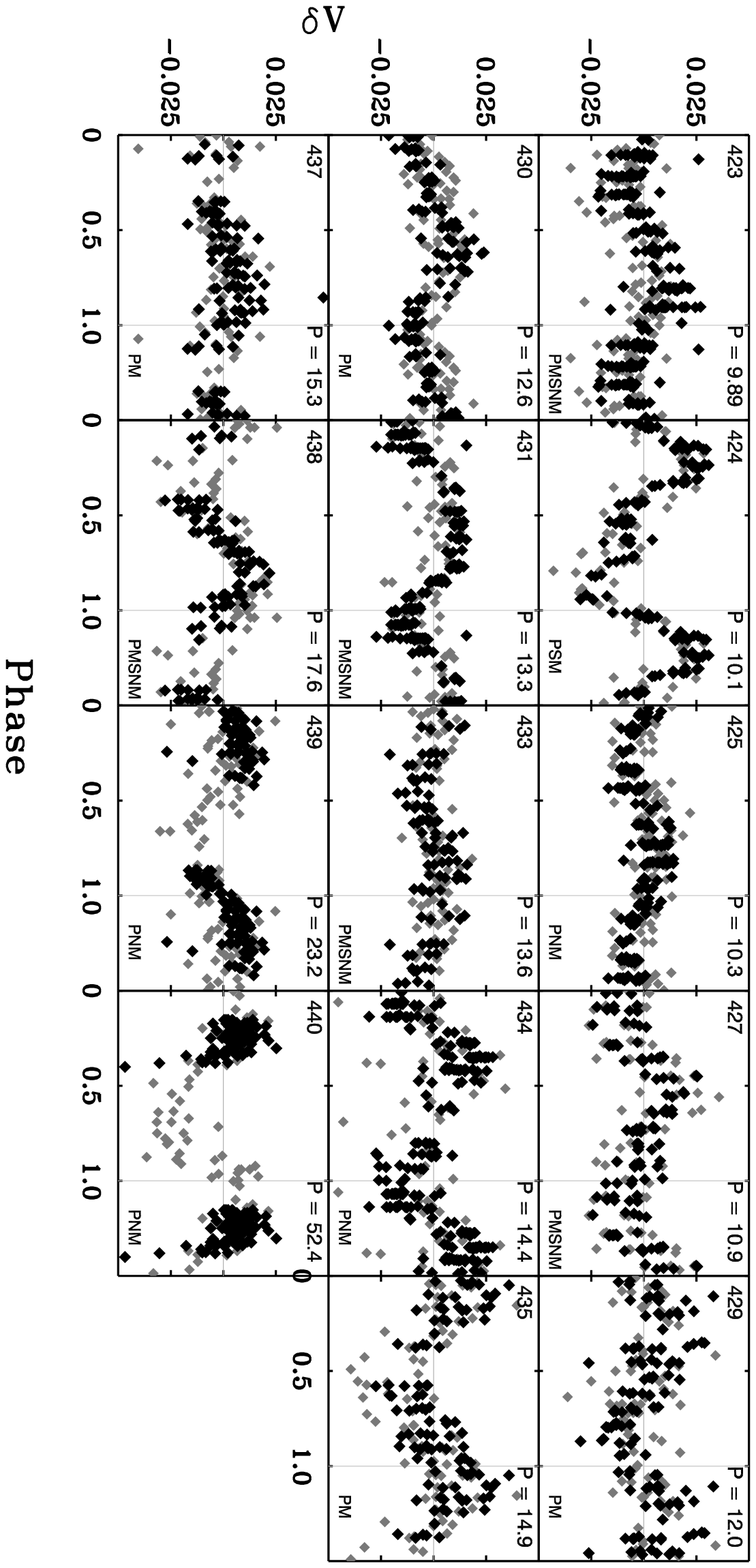}}\\
\centerline{Fig. 14. --- Continued.}
\clearpage
\pagestyle{plaintop}

\clearpage

% ***************************************************************************

\section{DATA FOR THE 441 STARS WITH MEASURED ROTATION PERIODS IN THE
FIELD OF M35}

Table~\ref{tab2} presents the results from this study together with
information relevant to this paper for 441 stars in the field of M35.
In the printed journal a stub version of Table 2 show the form of the
full table and a sample the first 5 lines of its contents. The full
version of the table can be found online. The stars appear in order
of increasing rotation period, and
the running number in the first column corresponds to the number in
the upper left hand corner of the stars light curve in Appendix A.
Columns 2 and 3 give the stellar equatorial coordinates (equinox 2000).
Column 4 lists the measured stellar rotation period in decimal days.
Columns 5, 6, and 7 gives the stellar V magnitude and B-V and V-I color
indices, respectively, corrected for extinction and reddening. Column
8 presents the number of radial-velocity measurements for the star and
columns 9 and 10 give the mean radial-velocity and the velocity standard
deviation, respectively. Column 11 list the radial-velocity cluster
membership probability calculated using the formalism by \citet{vkp58}.
Column 12 contains a proper-motion cluster membership probability from
either \citet{cudworth71} or \citet{ms86}. In column 13 we give the
abbreviated membership codes (initialisms) also found in the light
curves in Appendix A. The codes denote the type of membership information
available for the star and have the following meaning: Photometric
Member (PM; described in Section~\ref{phot}), Photometric Non-Member
(PNM), Photometric and Spectroscopic Member (PSM), Photometric and
Proper-motion Member (PPM), and Photometric Member but Spectroscopic
Non-Member (PMSNM). In column 14 we give the weights used for each star
when fitting the rotational isochrones in Section~\ref{isochrones} and
Section~\ref{mass-rotation}. Finally, in column 15 the rotational state
of the star is indicated by a 1-letter code representing, respectively,
the I sequence (``i''), the C sequence (``c''), and the gap (``g'').
Stars with a ``-'' in column 15 have locations in the color-period diagram
that do not correspond to either of the sequences or the gap.

\notetoeditor{An example of Table 2 with the first 5 lines can go here.
tab2.tex contains just the first 5 lines of data. tab2_online.tex is
the full table. NOTE: in the "b", and "c" table notes, the first letter
is missing.}

\clearpage
\input{tab2.tex}
\clearpage

% ***************************************************************************

\section{THE ROTATION PERIOD DISTRIBUTION OF NON-MEMBERS} \label{nm}

\begin{figure}[ht!]
\epsscale{1.0}
\plotone{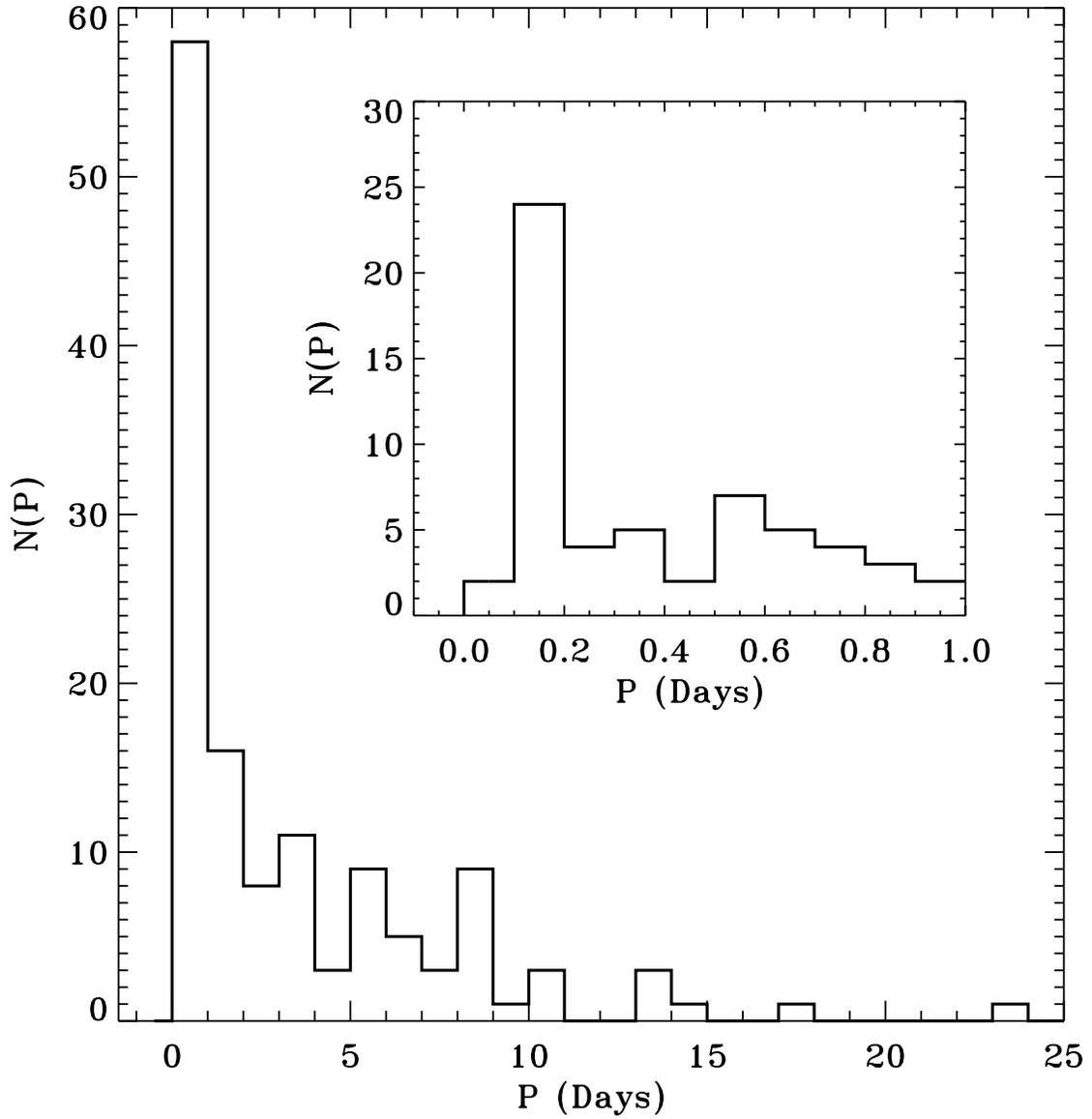}
\caption{The rotation period distribution of the 131 radial-velocity
and/or photometric non-members. The insert shows the distribution of
periods less than 1 day with increased resolution of 0.1 day.
\label{nmd}}
\end{figure}

In this section, we display and briefly comment upon the rotation
period distribution of the non-members among the 441 rotators, which
presumably are 
mostly field stars belonging to the Galactic disk. Unlike
the cluster members we do not know the ages, distances, or masses
of these stars. We will therefore only comment on a comparison
between the two distributions and on distinct features in the
period distribution of the 131 non-members shown in Figure~\ref{nmd}.
First, rotation periods are detected over approximately the same
range ($\sim$0.1 - 15 days) as for the 150 Myr cluster members,
with only a few stars rotating slower. Second, there appears to
be no indication of a bimodal distribution, but rather a distribution
with a peak of ultra fast rotators and a long tail of periods
up to and beyond 15 days. Third, and most strikingly, as shown
by the 0.1 day resolution of the insert in Figure~\ref{nmd}, the
non-member distribution exhibits a very distinct peak between 0.1
and 0.2 days. The phased light curves for these stars (Appendix A)
shows that the majority of these stars have large and well defined
photometric variability with with peak-to-peak amplitudes of $0\fm1$
or higher. It is possible that most of these stars are contact binaries
of, e.g., the W UMa-type. Such systems typically have orbital periods
of order 0.2-0.5 days and occur with a frequency of $\sim$0.2\%
\citep[OGLE Variable Star Catalog;][]{rucinski97}. The frequency
of rapidly varying non-members found in the field of M35 is $\sim$20
out of 13700 or $\sim$0.15\% and thus in good agreement with that
found from the OGLE Catalog. Indeed, if the light curves for these
stars represent eclipses rather than spot-modulation, then the
measured periods must be doubled bringing them into the expected
range for W Uma systems.

\clearpage
\clearpage

% ***************************************************************************
%\bibliographystyle{apj}
%\bibliography{refs}

\end{document}

%% file: tab2.tex
\begin{deluxetable}{rccrrcccrrccccc}
\tabletypesize{\scriptsize}
\setlength{\tabcolsep}{0.8mm}
\tablecaption{Data for the 441 stars with measured rotation periods in the
field of M35 \label{tab2}}
\tablewidth{0pt}
\tablehead{
\colhead{No.}\tablenotemark{a} &
\colhead{RA} &
\colhead{DEC} &
\colhead{$P_{rot}$} &
\colhead{$V_{0}$} &
\colhead{$(B-V)_{0}$} &
\colhead{$(V-I)_{0}$} &
\colhead{$N_{RV}$} &
\colhead{$\bar{RV}$} &
\colhead{$\sigma_{RV}$} &
\colhead{$P_{RV}$} &
\colhead{$P_{PM}$} &
\colhead{mcode}\tablenotemark{b} &
\colhead{Weight} &
\colhead{Sequence}\tablenotemark{c} \\
\colhead{} &
\colhead{$h~~m~~s$} &
\colhead{$\degr~~'~~''$} &
\colhead{Days} &
\colhead{} &
\colhead{} &
\colhead{} &
\colhead{} &
\colhead{$km~s^{-1}$} &
\colhead{$km~s^{-1}$} &
\colhead{\%} &
\colhead{\%} &
\colhead{} &
\colhead{} &
\colhead{} 
}
\startdata
  1 & 6 08 27.136 & 24 21 16.131 &  0.088 & 15.20 & 0.42 & 0.50 & ... & ...   & ...   & ... &  0 & PNM   & 0.0  & --  \\
  2 & 6 08 32.183 & 24 15 33.227 &  0.096 & 15.01 & 0.47 & 0.64 & ... & ...   & ...   & ... &  0 & PNM   & 0.0  & --  \\
  3 & 6 08 43.198 & 24 30 01.147 &  0.107 & 15.90 & 0.45 & 0.64 & ... & ...   & ...   & ... &  0 & PNM   & 0.0  & --  \\
  4 & 6 08 43.282 & 24 37 24.589 &  0.112 & 14.89 & 0.45 & 0.61 & ... & ...   & ...   & ... &  0 & PNM   & 0.0  & --  \\
  5 & 6 10 18.959 & 24 03 17.857 &  0.126 & 16.74 & 0.73 & 1.05 & ... & ...   & ...   & ... &  0 & PNM   & 0.0  & c   \\
  6 & 6 08 23.684 & 24 05 52.352 &  0.127 & 12.29 & 0.34 & 0.45 & 1   & 7.2   & 0.0   & 0   & 95 & PPM   & 0.75 & --  \\
  7 & 6 08 44.048 & 24 36 34.265 &  0.128 & 19.01 & ...  & 1.14 & ... & ...   & ...   & ... &  0 & PNM   & 0.0  & c   \\
  8 & 6 09 07.521 & 24 23 49.424 &  0.128 & 14.82 & 0.47 & 0.66 & ... & ...   & ...   & ... &  0 & PNM   & 0.0  & --  \\
\enddata

\tablenotetext{a}{Running number assigned to 310 rotators after sorting by rotation period.}
\tablenotetext{b}{Letter code denoting a star's cluster membership status (see introductory text of Appendix for code meaning).}
\tablenotetext{c}{Letters ``i'', ``c'', and ``g'' mark stars on the I and C sequances or in the gap, respectively.}

\end{deluxetable}